\documentclass[aps,prb,twocolumn,unsortedaddress,amsmath,amssymb]{revtex4-1}

\usepackage{graphicx}% Include figure files
\usepackage{dcolumn}% Align table columns on decimal point
\usepackage{bm}% bold math
\usepackage{tablefootnote}
\usepackage{hyperref}
\hypersetup{
	colorlinks=true,
	linkcolor=blue,
	citecolor=blue,
	urlcolor=blue
	}

\newcommand{\cm}{cm$^{-1}$}

\newcolumntype{d}[1]{D{.}{.}{#1}}
	
\begin{document}

\title{Collisional excitation of NH($^{3}\Sigma^{-}$) by Ar: a new $ab$ $initio$ 3D Potential Energy Surface and scattering calculations}
\author{D. Prudenzano \\ email:~prudenzano@mpe.mpg.de}
\affiliation{Center for Astrochemical Studies, 
           Max-Planck-Institut f\"ur extraterrestrische Physik,
           Gie\ss enbachstra\ss e 1, 85748 Garching, Germany}
\author{F. Lique}
\affiliation{LOMC - UMR 6294, CNRS-Universit\'e du Havre, 25 rue Philippe Lebon, BP 1123, 76063, Le Havre, France}
\author{R. Ramachandran}
\affiliation{LOMC - UMR 6294, CNRS-Universit\'e du Havre, 25 rue Philippe Lebon, BP 1123, 76063, Le Havre, France}
\author{L. Bizzocchi}
\affiliation{Center for Astrochemical Studies, 
           Max-Planck-Institut f\"ur extraterrestrische Physik,
           Gie\ss enbachstra\ss e 1, 85748 Garching, Germany}
\author{P. Caselli}
\affiliation{Center for Astrochemical Studies, 
           Max-Planck-Institut f\"ur extraterrestrische Physik,
           Gie\ss enbachstra\ss e 1, 85748 Garching, Germany}

\date{\today}% It is always \today, today,
             %  but any date may be explicitly specified

\begin{abstract}
Collisional excitation of light hydrides is important to fully understand the complex chemical and physical processes  of atmospheric and astrophysical environments. Here, we focus on the \mbox{NH(X$^3\Sigma^-$)-Ar} van der Waals system. First we have calculated a new three-dimensional Potential Energy Surface (PES), which explicitly includes the NH bond vibration. We have carried out the \textit{ab initio} calculations of the PES employing the open-shell single- and double-excitation couple cluster method with non-iterative perturbational treatment of the triple excitations [RCCSD(T)]. To achieve a better accuracy, we have first obtained the energies using the augmented correlation-consistent aug-cc-pVXZ (X = T, Q, 5) basis sets and then, we have extrapolated the final values to the complete basis set limit. We have also studied the collisional excitation of \mbox{NH(X$^3\Sigma^-$)-Ar} at the close-coupling level, employing our new PES. We calculated collisional excitation cross sections of the fine-structure levels of NH by Ar for energies up to 3000 \cm . After thermal average of the cross sections we have then obtained the rate coefficients for temperatures up to 350 K. The propensity rules between the fine-structure levels are in good agreement with those of similar collisional systems, even though they are not as strong and pronounced as for lighter systems, such as NH--He. The final theoretical values are also compared with the few available experimental data.     
\end{abstract}

%\pacs{Valid PACS appear here}% PACS, the Physics and Astronomy
                             % Classification Scheme.
%\keywords{Suggested keywords}%Use showkeys class option if keyword
                              %display desired
\maketitle

\section{\label{sec:1} Introduction}

The study of inelastic collisions plays a relevant role in the understanding of important processes in different fields, such as atmospheric and astrophysical chemistry and physics. In particular open-shell molecules are crucial, being highly reactive compounds and intermediate in a large number of chemical reactions.
A relevant chemical species is the NH radical. This compound serves as a prototype for other collisional studies involving open-shell molecules. Being diatomic it is also preferred for both experimental and theoretical scattering studies, owing to its large rotational energy level spacings.
In addition, the magnetic moment of its $^{3}\Sigma^{-}$ electronic ground state, makes NH suitable for studies of ultracold molecules \cite{friedrich2009, egorov2004}, because it can be easily thermalized at low temperatures through collision with cold buffer gas atoms. 
In the past, NH has been subject of many theoretical and experimental collisional studies in different electronic states and with a variety of perturbers, such as the rare gases He \cite{alexander1991quantum, rinnenthal2002state, krems2003low, cybulski2005interaction, stoecklin2009combining, tobola2011calculations, dumouchel2012fine, ramachandran2018new} and Ne \cite{rinnenthal2000state, kerenskaya2005experimental, bouhafs2015}.

In our work we focus on the calculation of a new \textit{ab initio} 3D-averaged Potential Energy Surface (PES) and collisional excitation for the NH($^{3}\Sigma^{-}$)-Ar system. To our knowledge, there are no theoretical scattering studies for the fine-structure excitation of NH($^{3}\Sigma^{-}$) by Ar, while there is only one experimental work performed by \citealt{dagdigian1989}, employing a crossed beam apparatus. However, this experiment provides only relative collisional cross sections up to the rotational level $N$=4 and no rate coefficients are available.

The most recent PES is given by \citealt{kendall1998}. They employed a combination of supermolecular and intermolecular unrestricted M{\o}ller-Plesset perturbation theory (UMPPT) \cite{cybulski1995partitioning,cybulski1989trurl} and a selection of monomer-centered basis sets augmented with bond functions. However, the NH bond length was kept frozen at 1.96 bohr. Recent studies \cite{kalugina2014new, bouhafs2015, lique2015, ramachandran2018new} have proven that the use of a 3D PES which takes into account molecular vibration leads to more accurate results when employed in collisional excitation studies of light hydrides by rare gases. Moreover, inclusion of the bond vibrational motion makes it possible to comprise excited vibrational states. Hence, we have computed a new \textit{ab initio} PES for the NH($^{3}\Sigma^{-}$)--Ar van der Waals complex including the NH bond vibration.

Then, we present the first fully quantum close-coupling (CC) calculations of rotational inelastic cross sections for the NH($^{3}\Sigma^{-}$)--Ar collisional system. In addition, we have taken into account the spin-coupling splitting of the rotational levels and we have included the temperature dependence of the fine-structure resolved rate coefficients in the final results. 

The paper is organized as follows: Sec. II covers the calculation of the new NH--Ar PES and informations about the bound states of the NH--Ar complex; in Sec. III, we present the scattering calculations, including the inelastic cross sections and rate coefficients. In Sec. IV, we compare the resulting cross sections with the available experimental data in Ref. \citenum{dagdigian1989}. Conclusions are given in Sec. V.

\section{\label{sec:2} Potential energy surface}

\begin{figure}
 \begin{center}
   \includegraphics[width=8.0cm,angle=0.]{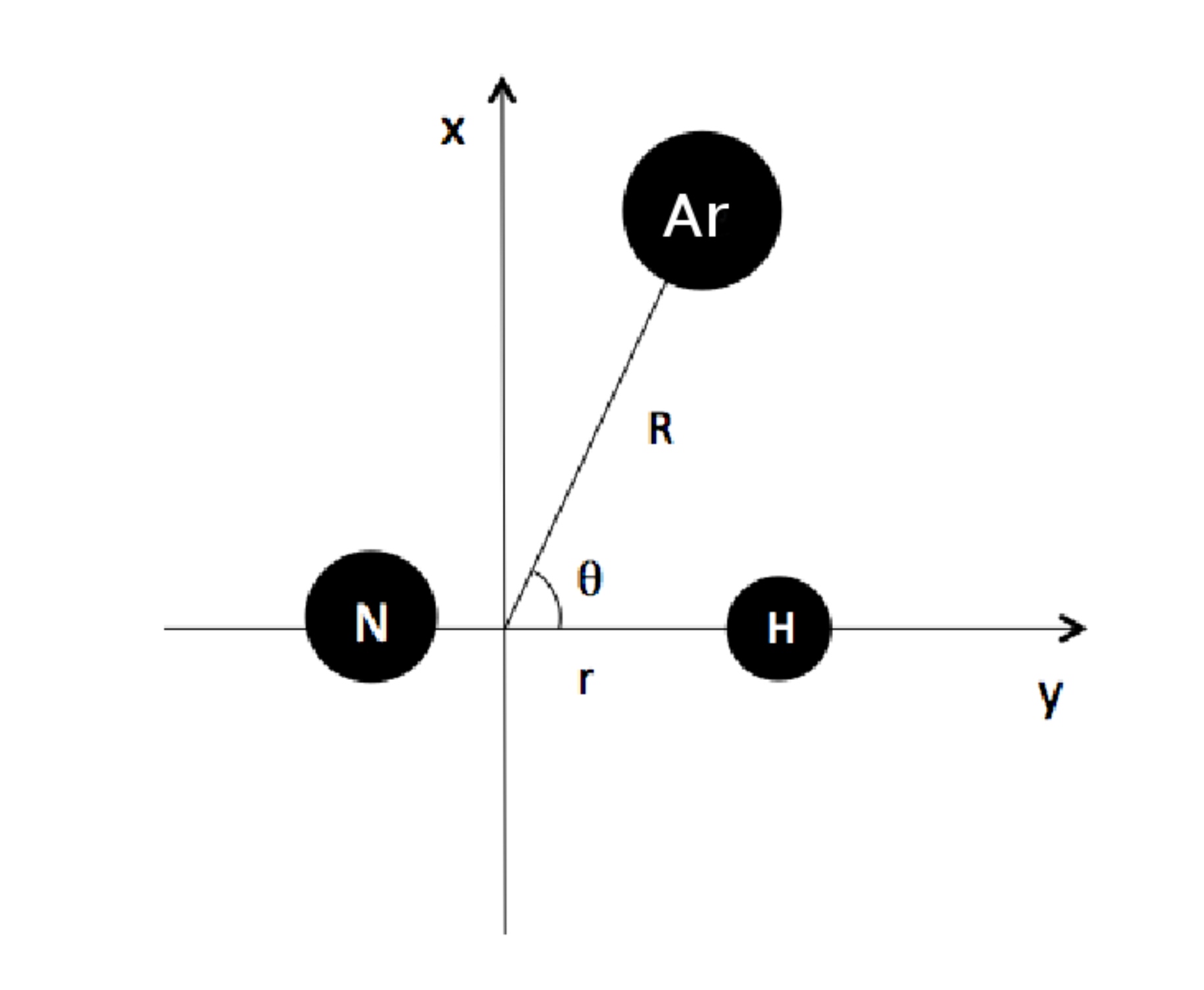}
   \caption{Definition of the Jacobi coordinate system. The origin of the coordinate system corresponds with the NH center of mass. R is the distance between the origin and the Ar atom, $\theta$ is the angle at which the Ar approach the NH molecule and r is the NH bond length.}
  \label{fig1}
 \end{center}
\end{figure}

The two interacting species are considered in their ground electronic states NH($^{3}\Sigma^{-}$) and Ar($^1S$). The NH($^{3}\Sigma^{-}$)--Ar van der Waals system has $^{3}$A$^{\prime\prime}$ ground electronic state. In this work, we used the Jacobi coordinate system (see Fig. \ref{fig1}). The center of coordinates is placed in the NH center of mass (c.m.), and the vector $\textbf{R}$ connects the NH c.m. with the Ar atom. The rotation of NH molecule is defined by the $\theta$ angle and the $r$ coordinate describes the NH bond length. 

We performed the calculations for five NH bond lengths $r$~=~[1.6, 1.8, 1.95, 2.15, 2.5]~bohr  which allows us to take into account vibrational motion of NH molecule up to $v = 2$.  We have carried out \emph{ab initio} calculations of the PES of the NH--Ar van der Waals complex at the partially spin-restricted coupled cluster with single, double and perturbative triple excitations [RCCSD(T)] \cite{Hampel92,watts1993coupled} level of theory, using MOLPRO 2015 package \citep{molpro10long}. In order to determine the interaction potential, $V(R,\theta, r)$, the basis set superposition error (BSSE) was corrected at all geometries using the Boys and Bernardi counterpoise scheme~\cite{boys1970calculation}:
\begin{eqnarray}
V(R,\theta, r) &=& E_{{\rm NH-Ar}}(R,\theta,r) \nonumber \\
& & -E_{{\rm NH}}(R,\theta,r)-E_{{\rm Ar}}(R,\theta,r)
\end{eqnarray}
\noindent where the energies of the NH and Ar monomers are computed using the full basis set of the complex.

To achieve a good description of the charge-overlap effects we have performed the
calculations in a rather large augmented correlation-consistent basis sets aug-cc-pVXZ (X = T, Q, 5) \cite{dunning1989gaussian}. Then, we have extrapolated the energies to the Complete Basis Set (CBS) limit using the following scheme \cite{peterson1994benchmark}:
\begin{equation}
\label{VCBS}
E_{\rm X}=E_{\rm CBS}+Ae^{-(X-1)}+Be^{-(X-1)^2},
\end{equation}
where $X$ is the cardinal number of the aug-cc-pVXZ basis set, $E_{\rm X}$ is the energy corresponding to aug-cc-pVXZ basis set, $E_{\rm CBS}$ is the energy extrapolated to CBS limit, $A$ and $B$ are the parameters to adjust. We have carried out the calculations for $\theta$ angle values from $0^{\circ}$ to $180^{\circ}$ in steps of $10^{\circ}$. $R$-distances were varied from 3.0 to 40.0\, bohr, yielding 52 points for each angular orientation.  Overall $\sim$5000 single point energies were calculated for the NH--Ar complex.

\subsection{Analytical representation of the potential energy surface}

The analytical expression employed for the interaction potential $V(R,\theta,r)$ has the following form \cite{werner1988adiabatic}:
\begin{equation}
\label{Vexp} V(R, \theta, r)= \sum_{n=1}^{N}\sum_{l=1}^{L}
B_{l,n}(R)(r-r_e)^{n-1}d_{m0}^{l+m-1} (\cos(\theta)),
\end{equation}
where
\begin{eqnarray}
\label{Bln}
B_{l,n}(R)=e^{-a_{l,n}(R-R_{l,n}^{(0)})}\Big(\sum_{i=0}^{2}b_{l,n}^{(i)}R^{i}\Big)\nonumber\\
-\frac{1}{2}\Big(1+\tanh\frac{R-R_{l,n}^{(1)}}{R_{l,n}^{ref}}\Big)\sum_{j=6,8,10}\frac{c_{l,n}^{(j)}}{R^j}.
\end{eqnarray}
\noindent The basis functions $d_{m0}^{l+m-1} (\cos(\theta))$ are Wigner rotation functions, $N$ is the total number of $r$-distances, and $L$ is the total number of angles.  The analytic potential was found to reproduce the calculated energies quite well: the mean difference between the analytic fit and the \textit{ab initio}
 computed interaction energies is less than 2\% over the entire grid. 

Previous studies \cite{kalugina2014new} have shown that averaging of the PES over corresponding vibrational level $v$ leads to a better agreement with experimental results than using a purely two-dimension PES. The newly constructed PES, which takes into account the stretching of the NH molecule, can be averaged over any vibrational state, up to $v$ = 2. The averaging is done using the following formula:
\begin{equation}
V_{v}(R, \theta) = \langle v(r) | V(R , \theta, r) | v(r) \rangle
\label{Eq:Vvib}
\end{equation}
The NH vibrational wave functions $| v(r) \rangle$ were those computed in Bouhafs et al. \cite{bouhafs2015} that were evaluated using a discrete variable representation (DVR) method \cite{colbert1992novel} from \textit{ab initio} calculations of the NH potential function using the internally contracted multireference configuration interaction (MRCI)\cite{werner1988efficient} level and a large aug-cc-pV5Z atomic basis set.  

\begin{figure}
\begin{center}
\includegraphics[width=8.0cm,angle=0.]{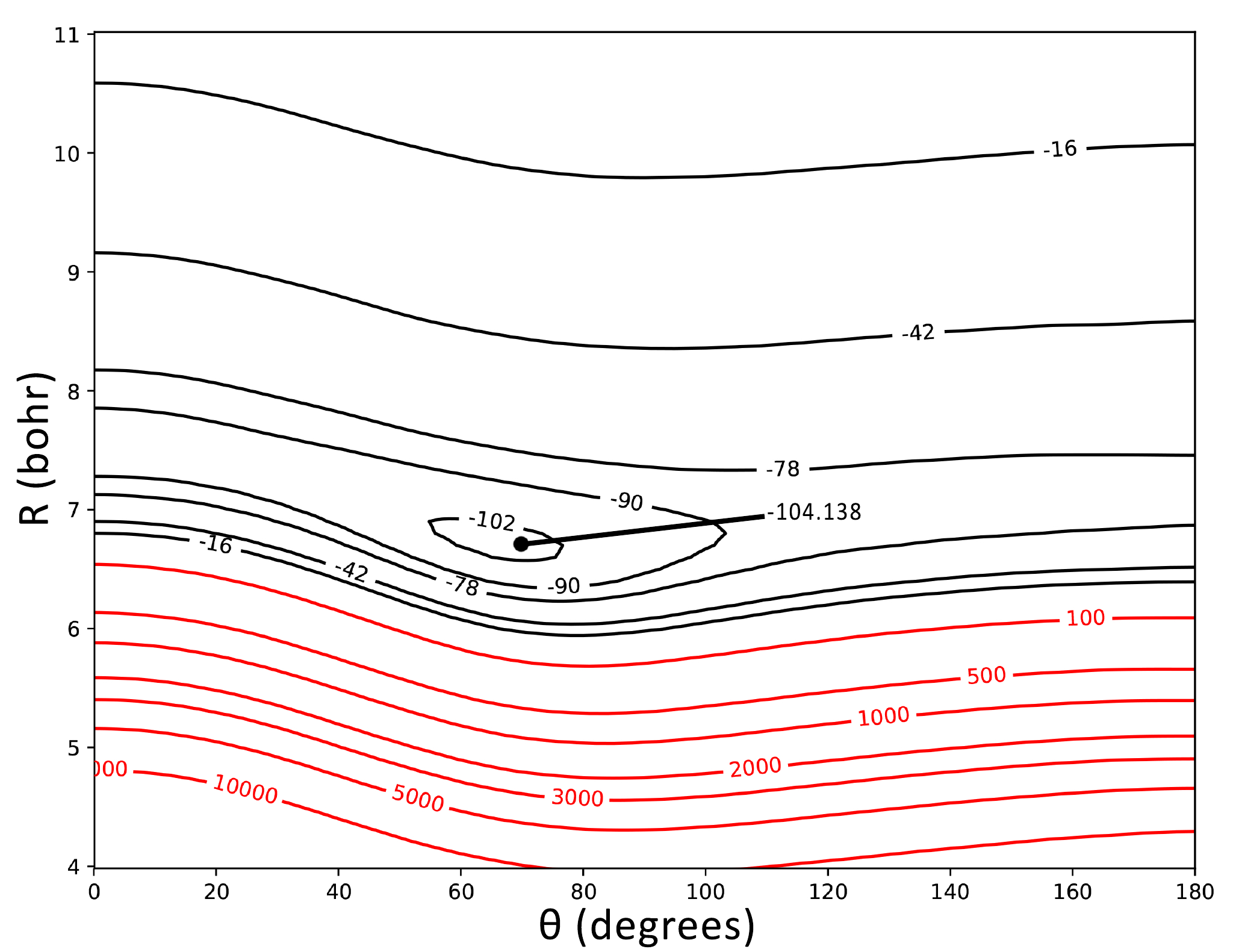}
\caption{Contour plot (in cm$^{-1}$) of the NH--Ar PES averaged over the ground vibrational state $v = 0$  as a function of Jacobi coordinates $R$ and $\theta$.} \label{fig2} 
\end{center}
\end{figure}

Figure \ref{fig2} depicts the contour plot of our 3D PES averaged over the ground vibrational state $v = 0$ as a function of $R$ and $\theta$ (hereafter refereed as 3D-ave PES). For this weakly-bound system the global minimum in the interaction energy was found to be -104.138 cm$^{-1}$ ($R=6.7$ bohr, $\theta=69^\circ$).

\begin{figure}
\begin{center}
\includegraphics[width=8.0cm,angle=0.]{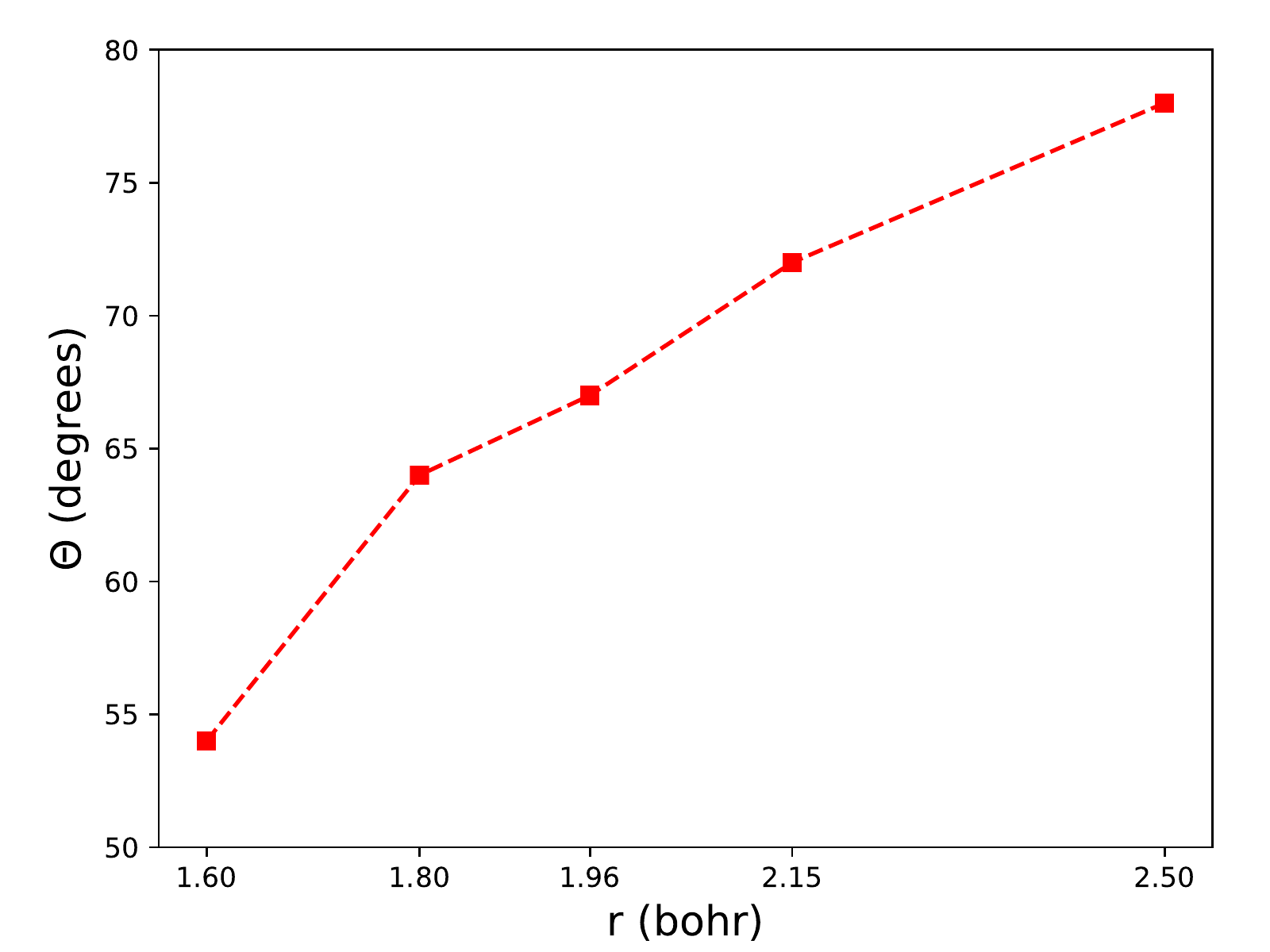}
\caption{Equilibrium angles at different NH bond lengths. The $R$ coordinate at the minimum energy does not change with $r$ and has always a value of $\sim$6.7 bohr.}\label{eq_angles} 
\end{center}
\end{figure}

Our study is in good agreement with the NH--Ar PES previously published \cite{kendall1998}. \citealt{kendall1998} carried out calculations for the NH--Ar interaction with the supermolecular unrestricted M{\o}ller-Plesset (UMP) perturbation theory and a combination of different basis sets. The NH intermolecular distance was fixed at 1.95 bohr. According to the authors the best results have been obtained with the aug-cc-pVTZ(ext-b) basis set, augmented with bond functions, and the global minimum is found at $R=6.75$ bohr and $\theta=67^\circ$, with a well-depth of -100.3 cm$^{-1}$ and an uncertainty within the 5{\%}. These values are very close to our results for $r = 1.95$ bohr ($R=6.7$, $\theta=67^\circ$, 103.787 cm$^{-1}$). Furthermore, the results of our 3D-ave PES also agree well with those listed above, confirming the high accuracy of our study. 
The slightly increased deepness of our well-depth is mostly due to the use of CBS extrapolation, since the energy follows a monotonic trend towards negative values, by approaching the infinite basis set limit.
Figure \ref{eq_angles} depicts the variation of the angle at which occurs the minimum of the interaction potential, for different NH bond distances. While the equilibrium angle changes substantially over increasing $r$, the $R$ distance is always close to $R=6.7$.

\subsection{\label{sec:3} NH--Ar bound states and dissociation energy}
 
Using the highly correlated 3D-ave PES described in the previous section, we have computed the bound states of NH--Ar complex using a coupled-channel approach, as implemented in the BOUND program \cite{bound}. The bound state calculations were performed for the main $^{14}$N and $^{40}$Ar isotopes. 

As a first step we performed bound state calculations neglecting the NH fine structure (i.e. NH was considered as a closed shell molecule). The calculations were performed with a propagator step size of 0.01 bohr, and the other propagation parameters were taken as the default BOUND values. The rotational basis includes the rotational states with $N_{max}  \leq  10$. The bound energy levels of the NH--Ar complex computed with the 3D-ave PES are listed in Table \ref{tab:1}. From the present calculations, dissociation energy ($D_{0}$) of the complex is 73.15 cm$^{-1}$ which is slightly larger than the previously calculated value of Kendall et al. \cite{kendall1998} ($D_0 = 71.5$ cm$^{-1}$). The difference (1.65 cm$^{-1}$) can be mainly attributed to the difference between the two NH--Ar PESs used in the calculations. Indeed, the well depth of the 3D-ave PES considering the vibration motion is slightly deeper (by few cm$^{-1}$) than the rigid rotor one of Kendall et al. \cite{kendall1998} and this difference leads to a larger estimated value of the dissociation energy.  

In order to derive the rotational constant of the NH--Ar complex, we have fitted the energies of Table  \ref{tab:1} to the rigid rotor expression: $E_J = E_{0} + BJ(J + 1) - DJ^{2}(J + 1)^{2}$ where $J$ corresponds to the total angular momentum of the complex. We have obtained for the rotational and quartic centrifugal distortion constants, $B=0.1087$ cm$^{-1}$ and $D=0.000025$ cm$^{-1}$. Such estimates allow generating the energetic structure of the complex and are useful for the interpretation of future experimental spectra. As a comparison, our rotational constant is in good agreement with the value obtained by \citealt{jansen1993theoretical}, i.e. B=0.1007.

\begin{table}[b]
\centering
\caption{NH--Ar bound energy levels (in cm$^{-1}$) obtained excluding the NH fine structure. Energies are relative to the ground-state energy of NH. All the levels correspond to the approximate quantum numbers $N = 0$. $J$ and $l$ correspond to the total and orbital angular momentum of the complex, respectively.} \label{tab:1}
\begin{tabular}{ccc}
 
\hline \hline 
$J$ & 				$l$    & Energy (cm$^{-1}$)  \\ \hline

0                                                    & 0                                              & -73.1507 \\
1                                                    & 1                                              & -72.9324 \\
2                                                    & 2                                              & -72.4988                                        \\
3                                                    & 3                                              & -71.8479 \\ \hline \hline   
\end{tabular}
\end{table}

As previously mentioned, the NH molecule exhibits a fine structure because of the coupling between the rotational angular momentum and the electronic spin. The BOUND program was modified to include this fine structure of the NH molecule \cite{ramachandran2018new}. Table \ref{tab:2} presents the bound state energies for the first  total angular momentum $J$. The predicted bound energy levels indicate that the coupling of the electron spin to the rotational motion of the complex is very weak. As a consequence, energy levels of NH--Ar are very similar to those obtained by neglecting the fine structure, as already found for the NH--He complex \cite{ramachandran2018new}. The dissociation energy is thus not significantly impacted by the fine structure. 

\begin{table}
\centering
\renewcommand{\baselinestretch}{1}
\caption{NH--Ar bound energy levels (in cm$^{-1}$) obtained with the inclusion of the NH fine structure. Energies are relative to the ground-state energy of NH. All the levels correspond to the approximate quantum numbers $N = 0, F_1$. $J$ and $l$ correspond to the total and orbital angular momentum of the complex, respectively.}\label{tab:2}
\begin{tabular}{ccc}

\hline \hline
$J$ & 				$l$    & Energy (cm$^{-1}$)  \\ \hline
1                           & 0              & -73.1519                              \\                                                                             
0                           & 1              & -72.8964                             \\
1                           & 1              & -72.9507                  	       \\
2						    & 1				 & -72.9305							   \\
1						    & 2				 & -72.4804							   \\
2						    & 2				 & -72.5169						       \\
3						    & 2				 & -72.4947							   \\
2						    & 3 		     & -71.8333				               \\
3						    & 3				 & -71.8658						       \\
4                           & 3              & -71.8426                             \\ \hline \hline   
\end{tabular}
\end{table}

\section{\label{sec:4} Scattering calculations}

Rotational transitions in the NH($^{3}\Sigma^{-}$) electronic ground state show fine-structure splitting, due to spin-rotation coupling. The rotational wave function of NH for $j\geq 1$ in the intermediate coupling scheme can be written as \cite{gordy1984,lique2005}:
\begin{eqnarray} \label{F}
|F_{1}jm\rangle & = & 
\cos\alpha|N=j-1,Sjm\rangle \nonumber \\
 & & +\sin\alpha|N=j+1,Sjm\rangle \nonumber \\
|F_{2}jm\rangle & = & |N=j,Sjm\rangle\\
|F_{3}jm\rangle & = & 
-\sin\alpha|N=j-1,Sjm\rangle \nonumber \\
 & & +\cos\alpha|N=j+1,Sjm\rangle \nonumber
\end{eqnarray}
where $|N,Sjm\rangle$ denotes pure Hund's case (b) basis functions and the mixing angle $\alpha$ is obtained by diagonalisation of the molecular Hamiltonian.
In this relation corresponding to the Hund's case (b), the total molecular angular momentum $j$ is defined by:
{\begin{equation}
{\bf j}={\bf N}+{\bf S}
\end{equation}
}
where $\mathbf{N}$ and $\mathbf{S}$ are the nuclear rotational and the electronic spin angular momenta. In the pure case (b) limit, $\alpha \to 0$, the $F_{1}$ level corresponds to $N=j-1$ and the $F_{3}$ level to $N=j+1$. The levels in the spin multiplets are usually labeled by the nuclear rotational quantum number $N$ and the spectroscopic index $F_i$. This notation will be used hereafter.

Using the new 3D-ave PES, we have studied the collisional excitation of NH by Ar. The scattering calculations were performed for the main $^{14}$N and $^{40}$Ar isotopes. The detailed description of the Close-Coupling (CC) calculations that consider the fine structure levels of the colliders is given in Ref.\citenum{lique2005}. The quantal coupled equations have been solved in the intermediate coupling scheme using the MOLSCAT code\cite{molscat1994} modified to take into account the fine structure of the rotational energy levels.

We used a total energy grid with a variable
steps. For the energies below 500~cm$^{-1}$ the step was equal to
1~cm$^{-1}$, then, between 500 and 1000~cm$^{-1}$ it was increased to
2~cm$^{-1}$, and to 20 for the interval 1000-3000~cm$^{-1}$. Using this
energy grid, the resonances (shape and Feshbach) that usually appear in
the cross sections at low energies were correctly represented.
 
In order to ensure convergence of the inelastic cross
sections, it is necessary to include in the calculations several
energetically inaccessible (closed) levels.  At the largest energies
considered in this work, the NH rotational basis was extended
to $N=12$ to ensure convergence of the rotational cross sections between levels with $N < 8$.
One also needs to converge inelastic cross sections with respect to partial waves. The total angular momentum quantum number $J$ needed for the convergence was set up to 238 for the inelastic cross sections. 

Moreover, in MOLSCAT, it is necessary to adjust the propagator's parameters in order to ensure convergence of cross sections calculations. 
For all the energies, the minimum and maximum integration distances were $R_{min}=3.0~\text{bohr}$ and $R_{max}= 50~\text{bohr}$, respectively. 
The \verb>STEPS> parameter was adjusted for each value of energy in order to obtain a step length of the integrator sufficient to achieve the convergence. 
In our work, the value of the \verb>STEPS> parameter decreases with increasing energy, going from 50 to 7 and, therefore, constraining the R spacing below 0.1-0.2 bohr at all energies. The reduced mass of the NH--Ar system is $\mu=10.912$~u and the NH($^{3}\Sigma^{-}$) rotational and centrifugal distortion constants have been taken from Ref. \citenum{lewen2004doppler}.

\begin{figure}
\begin{center}
\includegraphics[width=8cm,angle=0.]{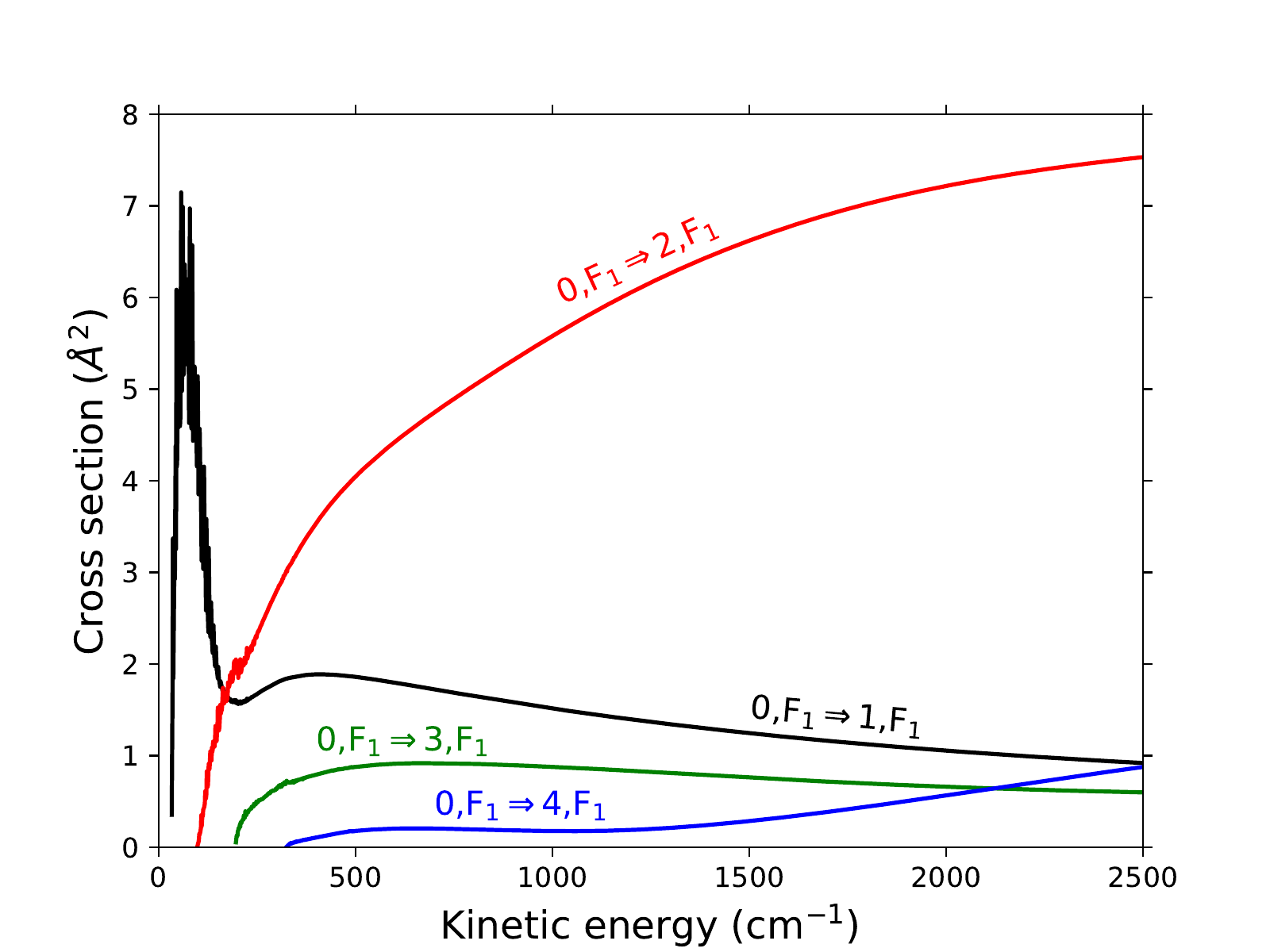}
\includegraphics[width=8cm,angle=0.]{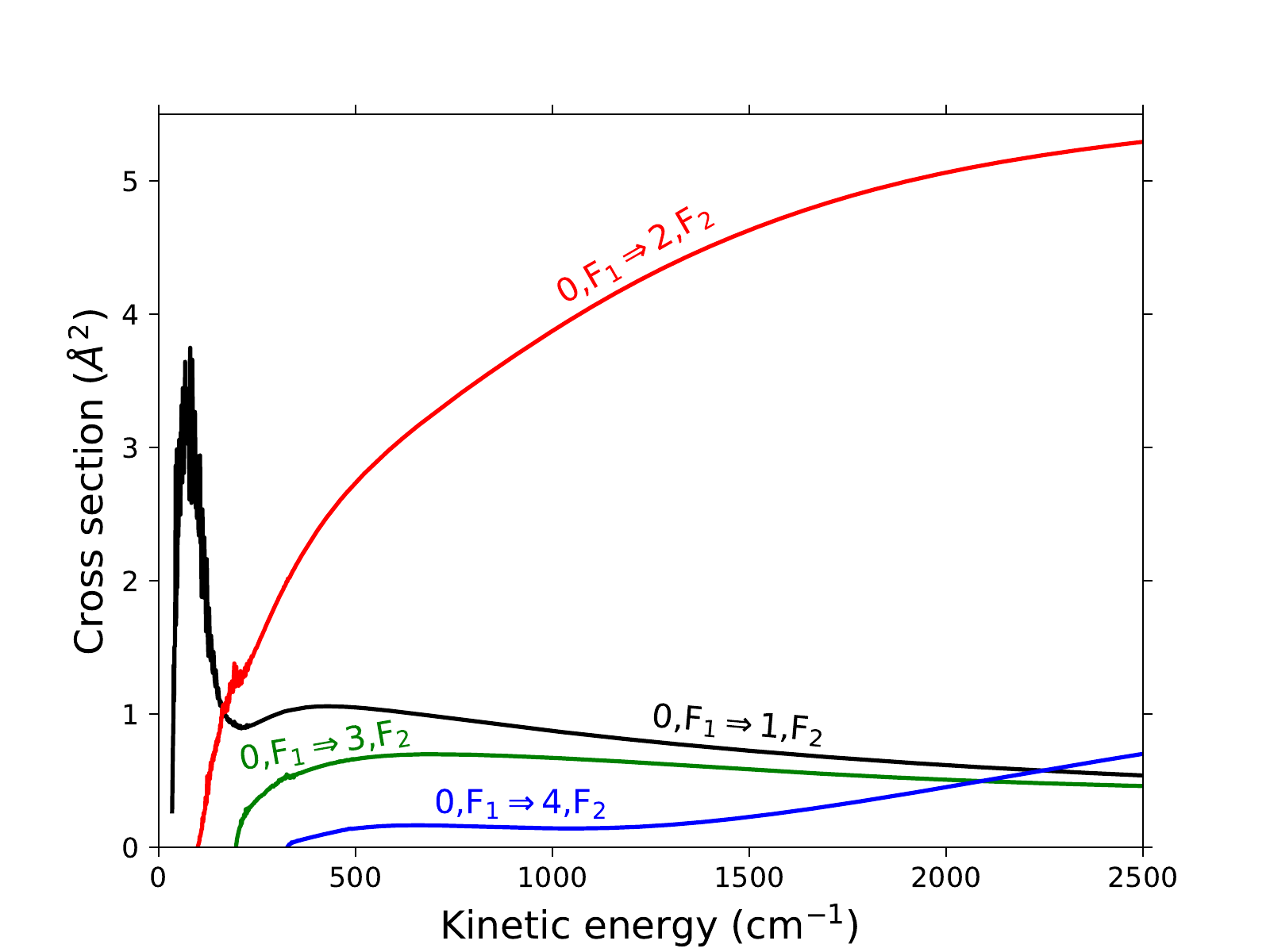}
\includegraphics[width=8cm,angle=0.]{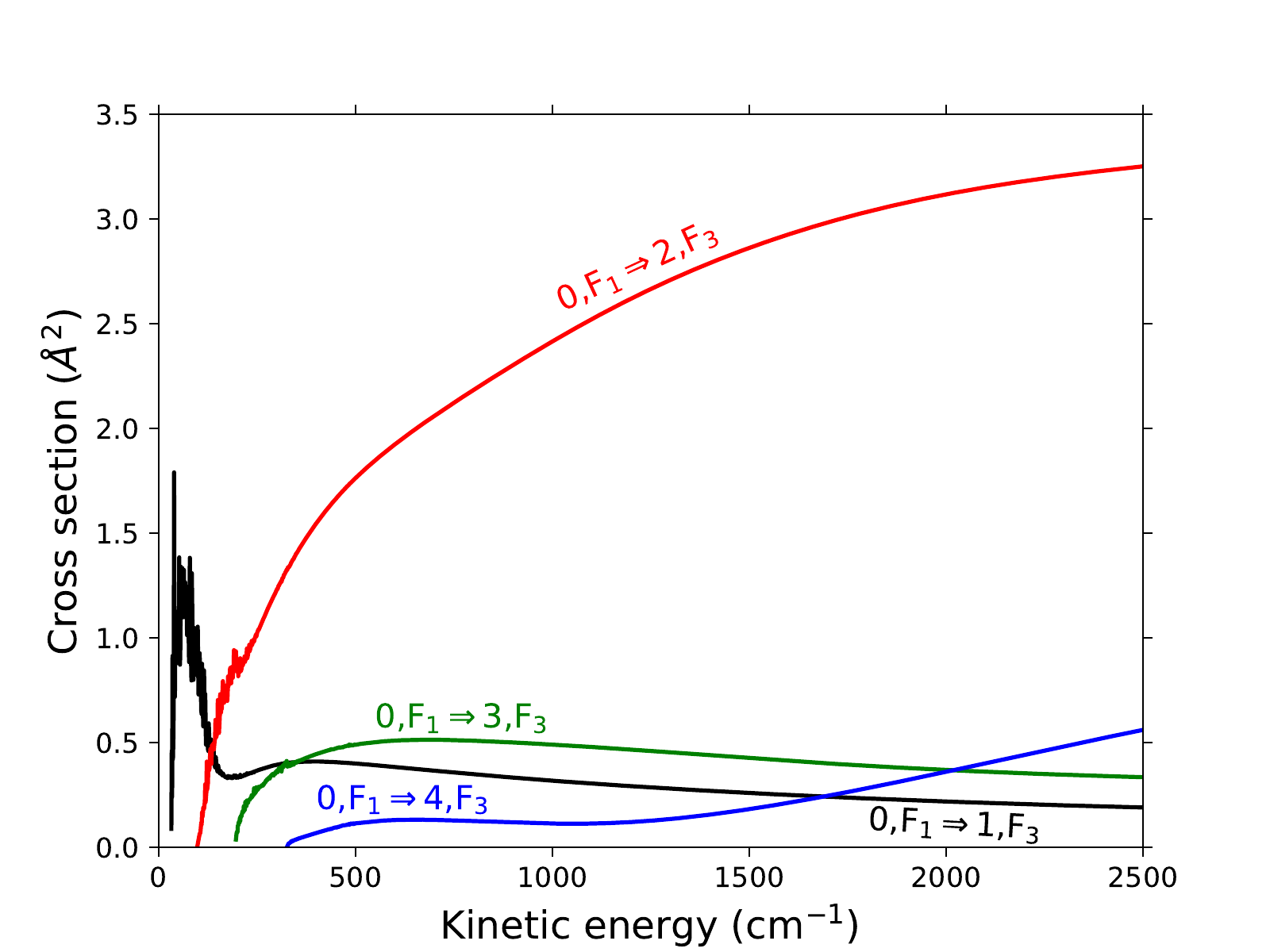}
\caption{Collisional excitation cross sections of NH by Ar from $N=0,F_{1}$. The upper panel is for fine-structure conserving transitions while the two other panels are for fine-structure changing transitions.}
\label{fig3}
\end{center}
\end{figure}

Figure~\ref{fig3} presents the energy variation of the integral cross sections for transitions from the initial rotational level $N=0,F_{1}$ of NH. 
The resonances shown at low collisional energies are related to the presence of a $\sim$104 cm$^{-1}$ deep attractive potential well. As a consequence, the Ar atom can be temporarily trapped there forming quasi-bond states before dissociation of the complex \cite{smith1979rot, christoffel1983complex}. However, excitation cross sections are less affected and therefore show few resonances. Indeed, the energy spacing between rotational levels is generally larger than the well depth of the PES.   

The magnitude of the cross sections shown in figure~\ref{fig3} seems to present the following propensity rules:

(1) Overall decreasing of the cross sections with increasing $\Delta N$, according to the usual trend for rotational excitation. In addition, even $\Delta N$ transitions are favored over odd $\Delta N$ transitions as a consequence of near-homonuclearity of the PES.

(2) Fine-structure conserving transitions are favored, i.e. $\Delta j = \Delta N$ in the case of pure Hund's case (b).   

The same propensity rules are shown in similar systems, such as NH-He and NH-Ne collisions \cite{tobola2011calculations, bouhafs2015, ramachandran2018new}. In particular, the latter applies in general to molecules in the $^{3}\Sigma^{-}$ electronic state. Both porpensity rules have been predicted theoretically \citep{alexander1983propensity} and also observed for the O$_{2}$-He \cite{orlikowski1985,lique2010temperature} or SO(X$^3\Sigma^-$)--He\cite{lique2005} collisions.   

The thermal rate coefficients, $k_{F_{i}j \to F_{i}'j'}(T)$, for excitation and de-excitation transitions between fine-structure levels of NH can be calculated by averaging CC excitation cross sections, $\sigma_{F_{i}j \to F_{i}'j}$, over a Maxwellian distribution of collision velocities, as follows:
\begin{eqnarray}
k_{F_{i}j \to F_{i}'j'}(T) & = & \left(\frac{8k_{B}T}{\pi\mu}\right)^{\frac{1}{2}}\left(\frac{1}{k_{B}T}\right)^{2}  \nonumber  \\
&  & \times  \int_{0}^{\infty}E_{k} \sigma_{F_{i}j \to F_{i}'j'} (E_{k}) e^{\frac{-E_{k}}{k_{B}T}}dE_{k}
\end{eqnarray}
where $k_{B}$ is the Boltzmann constant, $\mu$ is the reduced mass of the system and $E_k$ is the translational energy.\\

\begin{figure}
\begin{center}
\includegraphics[width=8cm,angle=0.]{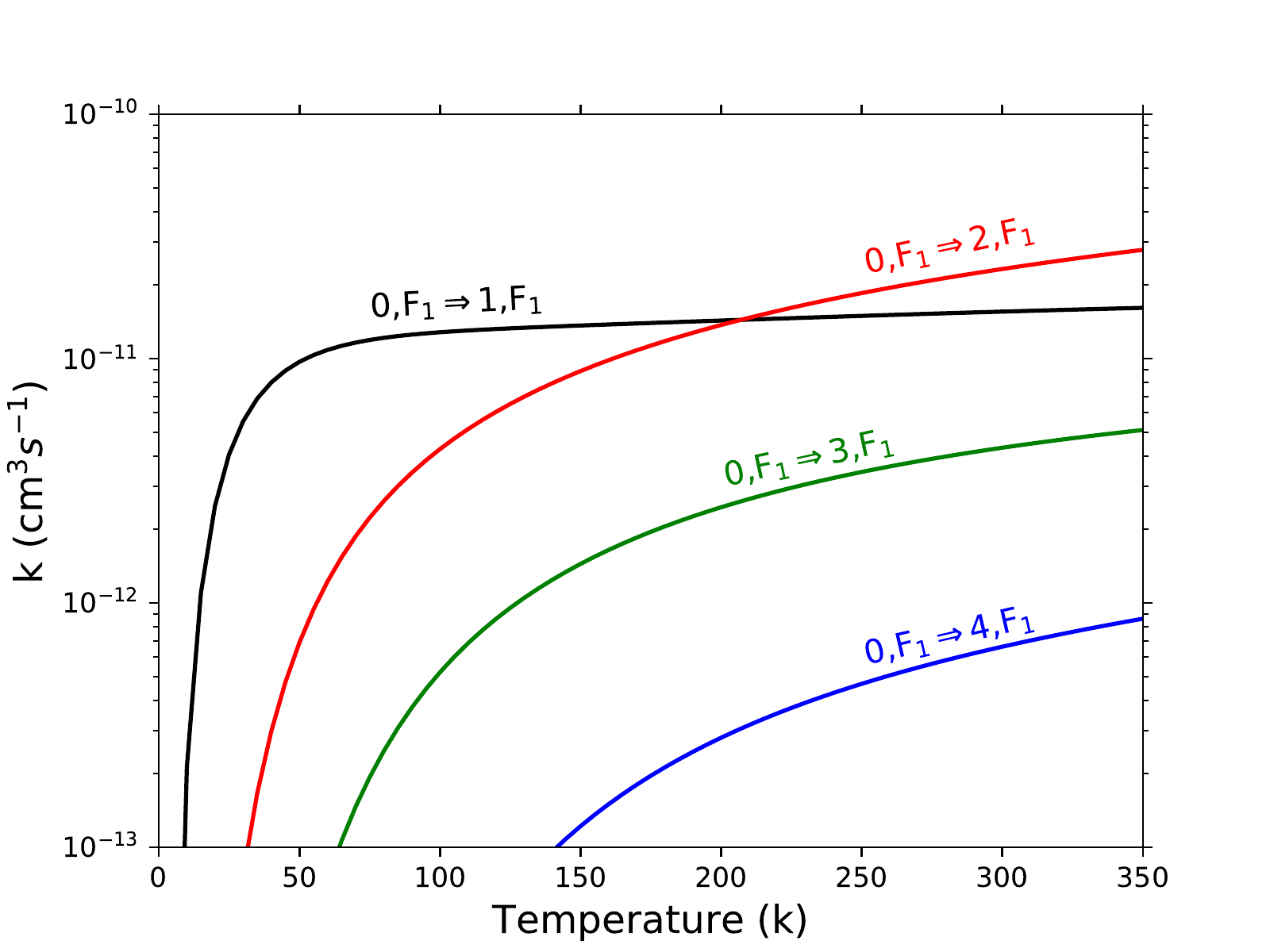}
\includegraphics[width=8cm,angle=0.]{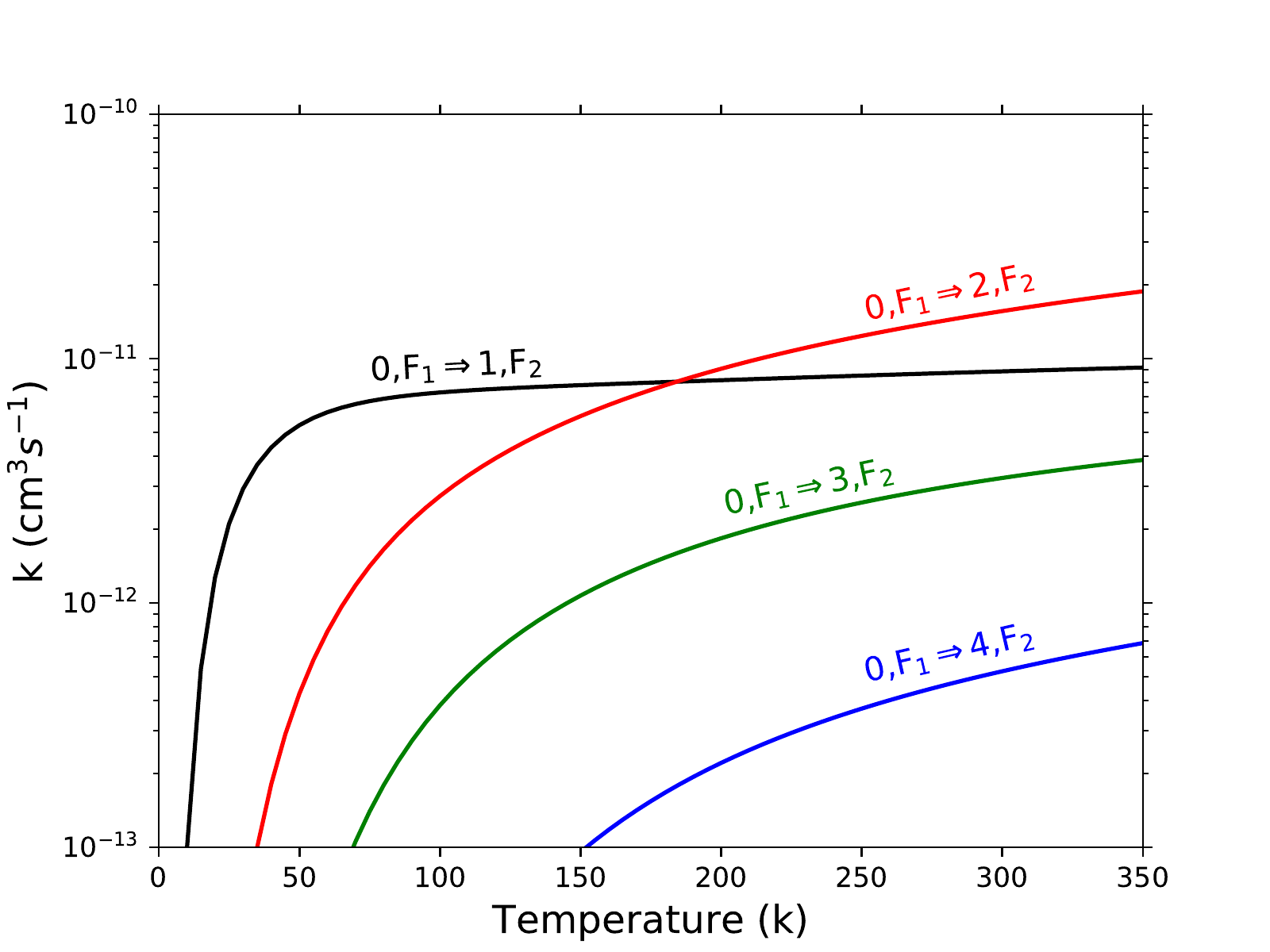}
\includegraphics[width=8cm,angle=0.]{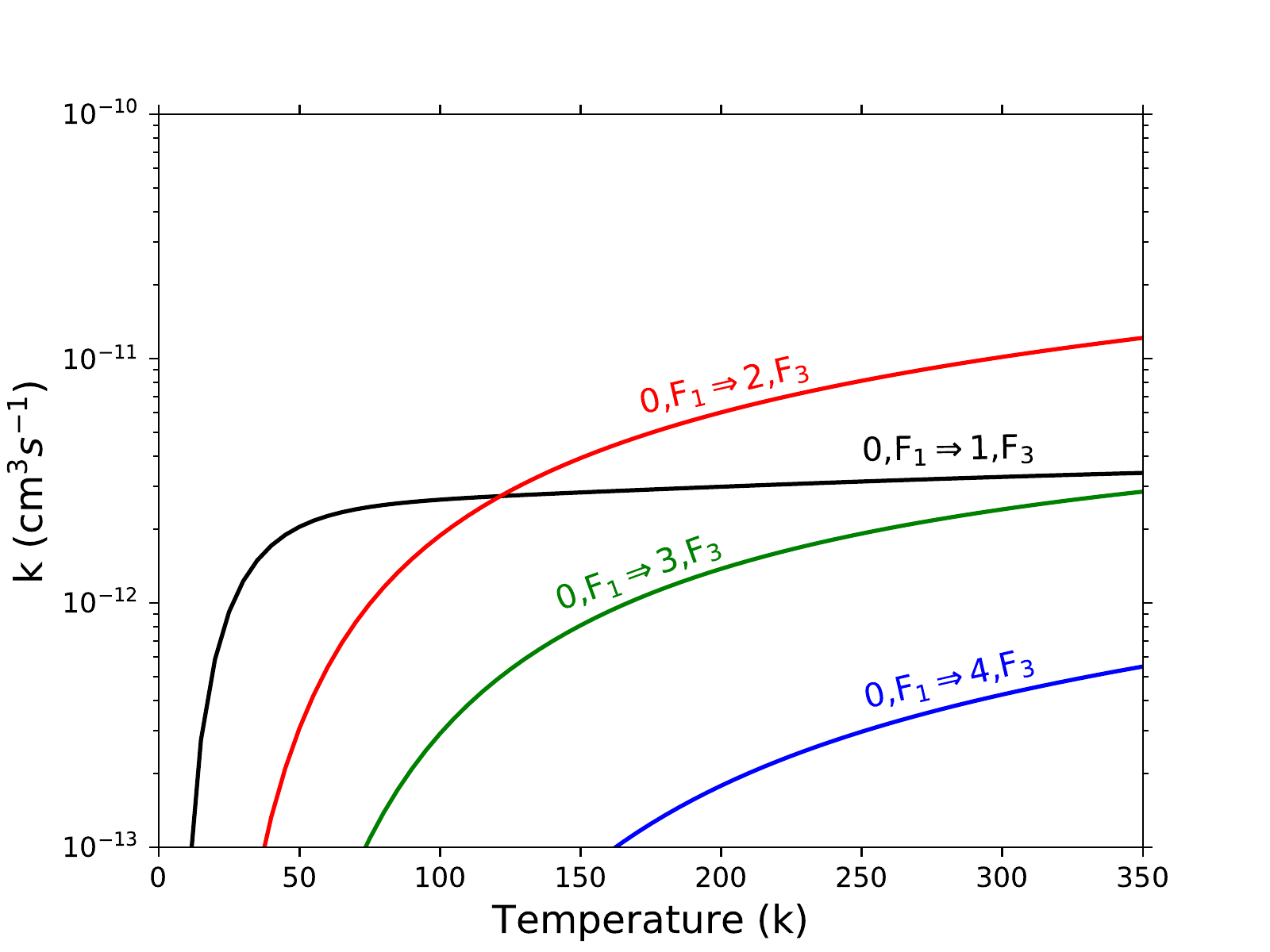}
\caption{Thermal dependence of the rate coefficients of NH by Ar from $N=0,F_{1}$. The upper panel is for fine-structure conserving transitions while the two other panels are for fine-structure changing transitions.}
\label{fig4}
\end{center}
\end{figure}

The thermal dependence of these state-to-state rate
coefficients for temperatures up to 350~K is shown in Fig.~\ref{fig4} for transitions out of the $N=0, j=1,F_1$ level.  

The rate coefficients display the same propensity rules as seen for the integral cross sections. In particular, the rate coefficients for $F$-conserving transitions are generally larger than those for $F$-changing transitions.

\section{Comparison with experiments}

\begin{figure}
\begin{center}
\includegraphics[width=8.5cm]{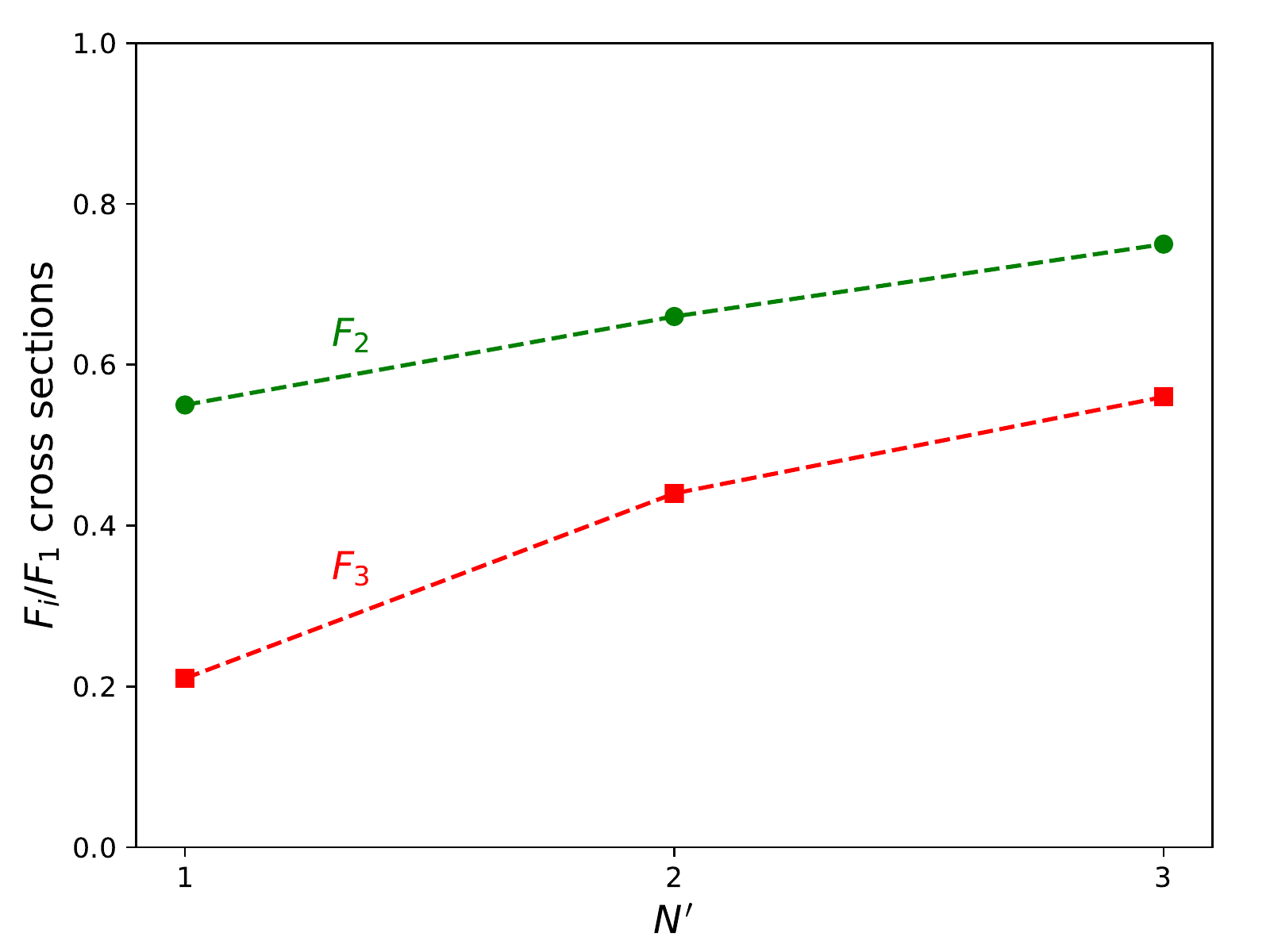}
\caption{Scaling relation for different $F$-levels over increasing final $N'$. The values correspond to the $N=0,F_1\rightarrow N'=X,F_2$ and $N=0,F_1\rightarrow N'=X,F_3$ cross sections normalized with respect to the $N=0,F_1\rightarrow N'=X,F_1$ one, with X = 1, 2 and 3.}
\label{comp_scaling}
\end{center}
\end{figure}

Our new calculated cross sections can be compared with the previous experimental results, obtained for a collisional energy of 410 cm$^{-1}$ and for rotational levels up to $N$=4,$F_1$ (Ref. \citenum{dagdigian1989}). 
\begin{table}[b]
\centering
  \caption{Comparison between experimental and our theoretical cross sections at a collisional energy of 410 cm$^{-1}$ and for transitions out of the $N=0,F_1$ rotational level. All the values are normalized with respect to the cross section for the $N=0,F_1\rightarrow N'=1, F_1$ transition. Experimental error in parenthesis are in units of the last quoted digit.}  
  \setlength{\tabcolsep}{2.2pt}
      \begin{tabular}{ccccccc}
\hline\hline
          & \multicolumn{2}{c}{$F_1$} & \multicolumn{2}{c}{$F_2$} & \multicolumn{2}{c}{$F_3$} \\
          \hline  
    \multicolumn{1}{c}{$N'$} & Exp\footnotemark[1]  & Theory\footnotemark[2]  & Exp\footnotemark[1]    & Theory\footnotemark[2]   & Exp\footnotemark[1]    & Theory\footnotemark[2] \\
    \hline
    \multicolumn{1}{c}{1} & 1.0 & 1.0   & 0.662(40) & 0.560 & 0.255(54) & 0.217 \\
    \multicolumn{1}{c}{2} & 0.407(54) & 1.906 & 0.284(46) & 1.275 & 0.154(20)   & 0.833 \\
    \multicolumn{1}{c}{3} & 0.068(15) & 0.427 & 0.059(10) & 0.321 & 0.047(08) & 0.239 \\
    \multicolumn{1}{c}{4} & 0.023(05)   &  0.061  &  -  & - & - & - \\
\hline\hline
     \end{tabular}
 \label{comp_exp}
   \footnotetext[1]{Ref. \citenum{dagdigian1989}}
   \footnotetext[2]{Our work.}
\end{table}
Tab. \ref{comp_exp} shows experimental and theoretical values normalized with respect to the $N=0,F_1\rightarrow N'=1, F_1$ cross section.
The $F$-conserving propensity rule is overall fullfilled in both the experimental and calculated values. In addition, the $F$-conserving cross sections follow the simple scaling relation observed by \citealt{dagdigian1989}, as shown in Fig.~\ref{comp_scaling}.
The main discrepancy is the trend of the cross sections over increasing $\Delta N$ and over even/odd $\Delta N$, as discussed for the first propensity rule in Sec. III. Furthermore, according to the results of Ref. \citenum{dagdigian1989}, the largest cross sections are those with $N'=1$, whereas this is not the case in our study. In fact, larger values are related to the transitions involving $N'=2$, as also shown in figure~\ref{fig3}.

It is likely that these discrepancies are due to a particular feature of the experiment. In fact, as declared by the author, the NH beam was not entirely pure, with 68\% of the population in the rotational ground state $N=0,F_1$, and approximately 16\% and 9\% in the $N=1,F_1$ and $N=1,F_2$ levels, respectively. 
By taking into account this NH beam population composition, the propensity rules observed in the experiment can be reproduced making a convolution of the various cross sections involved.
This is shown in Tab. \ref{comp_exp2}, which gathers values computed using 68\% contribution from inelastic cross section for transitions out of the $N=0,F_1$, 16\% from cross sections involving the $N=1,F_1$, and 9\% from those involving the $N=1,F_2$.

It should be pointed out that there is a 7\% population with unknown distribution and thus the theoretical results obtained through convolution are still different in magnitude from the experimental ones.

\begin{table*}[t]
\centering
  \caption{Comparison between experimental and convolved theoretical cross sections at a collisional energy of 410 cm$^{-1}$ and for transitions out of the $N=0,F_1$ rotational level. All the values are normalized with respect to the cross section of the $N=0,F_1\rightarrow N'=1, F_1$ transition. Experimental error in parenthesis are in units of the last quoted digit. The convolution of the theoretical values is described in Sec. IV. \label{comp_exp2}}  
  \setlength{\tabcolsep}{4pt}
      \begin{tabular}{cccccccccc}
\hline\hline
          & \multicolumn{3}{c}{$F_1$} & \multicolumn{3}{c}{$F_2$} & \multicolumn{3}{c}{$F_3$} \\
          \hline  
    \multicolumn{1}{c}{$N'$} & Exp\footnotemark[1]  & Theory\footnotemark[2] & Exp\footnotemark[1]  & Theory\footnotemark[2] & Exp\footnotemark[1]     & Theory\footnotemark[2] \\
    \hline
    \multicolumn{1}{c}{1} & 1.0 & 1.0 & 0.662(40)     & 0.842 & 0.255(54) & 0.365 \\
    \multicolumn{1}{c}{2} & 0.407(54) & 0.954  & 0.284(46)    & 0.654 & 0.154(20)   & 0.411 \\
    \multicolumn{1}{c}{3} & 0.068(15) & 0.224  & 0.059(10)    & 0.172 & 0.047(08) & 0.123 \\
    \multicolumn{1}{c}{4} & 0.023(05) & 0.037 & - & - & - & - \\
\hline\hline
     \end{tabular}
   \footnotetext[1]{Ref. \citenum{dagdigian1989}}
   \footnotetext[2]{Our work. See text for details.}
\end{table*}

\section{Conclusion}

We have computed a new highly accurate 3D PES for the NH--Ar collisional system by taking into account the stretching of the NH bond. We carried out these \textit{ab initio} calculations at the RCCSD(T) level and a complete basis set extrapolation. The results are in good agreement with the most recent PES available \cite{kendall1998}. 

Employing our new 3D-ave PES we have calculated the dissociation energy of the NH--Ar van der Waals complex and the corresponding rotational and centrifugal distortion constants. We have also performed scattering calculations at the close-coupling level, obtaining collisional cross sections for energies up to 3000 cm$^{-1}$. We have then determined rate coefficients for temperatures up to 350 K. The resulting values follows the same propensity rules seen in other similar systems \citep{bouhafs2015,ramachandran2018new}, i.e. overall decreasing with increasing $\Delta N$, even $\Delta N$ favored over odd $\Delta N$ and larger values for $F$-conserving transitions. 

Our theoretical results have been compared to a previous experimental study\cite{dagdigian1989}. The discrepancy concerning the $\Delta N$ propensity rules can be explained with the impurity of the NH population of the experimental molecular beam, since we have been able to reproduce the results of the experiment through convolution of various cross sections, as discussed in Sec. IV. 

We hope that our results will encourage new experimental studies concerning collisional excitation of NH($^{3}\Sigma^{-}$) by Ar. In particular it would be interesting to fill the gap of missing data regarding Ar as a collisional partner, with respect to systems involving He or Ne, more widely studied. Furthermore, a complete overview of these systems could also encourage studies with ortho- and para-H$_2$, highly important for astrophysical environments. 

\section*{Supplementary Material}
The supplementary material provides the analytic form of the NH--Ar potential energy surface and the NH--Ar collisional rate coefficients. 
 
\section*{Acknowledgements}
F. L. acknowledges the Institut Universitaire de France. We would also thank P. Dagdigian for the interesting discussion about the comparison between our theoretical and his experimental results.

\bibliography{vanderwaals}

%merlin.mbs apsrev4-1.bst 2010-07-25 4.21a (PWD, AO, DPC) hacked
%Control: key (0)
%Control: author (8) initials jnrlst
%Control: editor formatted (1) identically to author
%Control: production of article title (-1) disabled
%Control: page (0) single
%Control: year (1) truncated
%Control: production of eprint (0) enabled
\begin{thebibliography}{39}%
\makeatletter
\providecommand \@ifxundefined [1]{%
 \@ifx{#1\undefined}
}%
\providecommand \@ifnum [1]{%
 \ifnum #1\expandafter \@firstoftwo
 \else \expandafter \@secondoftwo
 \fi
}%
\providecommand \@ifx [1]{%
 \ifx #1\expandafter \@firstoftwo
 \else \expandafter \@secondoftwo
 \fi
}%
\providecommand \natexlab [1]{#1}%
\providecommand \enquote  [1]{``#1''}%
\providecommand \bibnamefont  [1]{#1}%
\providecommand \bibfnamefont [1]{#1}%
\providecommand \citenamefont [1]{#1}%
\providecommand \href@noop [0]{\@secondoftwo}%
\providecommand \href [0]{\begingroup \@sanitize@url \@href}%
\providecommand \@href[1]{\@@startlink{#1}\@@href}%
\providecommand \@@href[1]{\endgroup#1\@@endlink}%
\providecommand \@sanitize@url [0]{\catcode `\\12\catcode `\$12\catcode
  `\&12\catcode `\#12\catcode `\^12\catcode `\_12\catcode `\%12\relax}%
\providecommand \@@startlink[1]{}%
\providecommand \@@endlink[0]{}%
\providecommand \url  [0]{\begingroup\@sanitize@url \@url }%
\providecommand \@url [1]{\endgroup\@href {#1}{\urlprefix }}%
\providecommand \urlprefix  [0]{URL }%
\providecommand \Eprint [0]{\href }%
\providecommand \doibase [0]{http://dx.doi.org/}%
\providecommand \selectlanguage [0]{\@gobble}%
\providecommand \bibinfo  [0]{\@secondoftwo}%
\providecommand \bibfield  [0]{\@secondoftwo}%
\providecommand \translation [1]{[#1]}%
\providecommand \BibitemOpen [0]{}%
\providecommand \bibitemStop [0]{}%
\providecommand \bibitemNoStop [0]{.\EOS\space}%
\providecommand \EOS [0]{\spacefactor3000\relax}%
\providecommand \BibitemShut  [1]{\csname bibitem#1\endcsname}%
\let\auto@bib@innerbib\@empty
%</preamble>
\bibitem [{\citenamefont {Friedrich}\ and\ \citenamefont
  {Doyle}(2009)}]{friedrich2009}%
  \BibitemOpen
  \bibfield  {author} {\bibinfo {author} {\bibfnamefont {B.}~\bibnamefont
  {Friedrich}}\ and\ \bibinfo {author} {\bibfnamefont {J.~M.}\ \bibnamefont
  {Doyle}},\ }\href@noop {} {\bibfield  {journal} {\bibinfo  {journal}
  {ChemPhysChem}\ }\textbf {\bibinfo {volume} {10}},\ \bibinfo {pages} {604}
  (\bibinfo {year} {2009})}\BibitemShut {NoStop}%
\bibitem [{\citenamefont {Egorov}\ \emph {et~al.}(2004)\citenamefont {Egorov},
  \citenamefont {Campbell}, \citenamefont {Friedrich}, \citenamefont {Maxwell},
  \citenamefont {Tsikata}, \citenamefont {Van~Buuren},\ and\ \citenamefont
  {Doyle}}]{egorov2004}%
  \BibitemOpen
  \bibfield  {author} {\bibinfo {author} {\bibfnamefont {D.}~\bibnamefont
  {Egorov}}, \bibinfo {author} {\bibfnamefont {W.}~\bibnamefont {Campbell}},
  \bibinfo {author} {\bibfnamefont {B.}~\bibnamefont {Friedrich}}, \bibinfo
  {author} {\bibfnamefont {S.}~\bibnamefont {Maxwell}}, \bibinfo {author}
  {\bibfnamefont {E.}~\bibnamefont {Tsikata}}, \bibinfo {author} {\bibfnamefont
  {L.}~\bibnamefont {Van~Buuren}}, \ and\ \bibinfo {author} {\bibfnamefont
  {J.}~\bibnamefont {Doyle}},\ }\href@noop {} {\bibfield  {journal} {\bibinfo
  {journal} {Eur. Phys. J. D}\ }\textbf {\bibinfo {volume} {31}},\ \bibinfo
  {pages} {307} (\bibinfo {year} {2004})}\BibitemShut {NoStop}%
\bibitem [{\citenamefont {Alexander}\ \emph {et~al.}(1991)\citenamefont
  {Alexander}, \citenamefont {Dagdigian},\ and\ \citenamefont
  {Lemoine}}]{alexander1991quantum}%
  \BibitemOpen
  \bibfield  {author} {\bibinfo {author} {\bibfnamefont {M.~H.}\ \bibnamefont
  {Alexander}}, \bibinfo {author} {\bibfnamefont {P.~J.}\ \bibnamefont
  {Dagdigian}}, \ and\ \bibinfo {author} {\bibfnamefont {D.}~\bibnamefont
  {Lemoine}},\ }\href@noop {} {\bibfield  {journal} {\bibinfo  {journal} {J.
  Chem. Phys}\ }\textbf {\bibinfo {volume} {95}},\ \bibinfo {pages} {5036}
  (\bibinfo {year} {1991})}\BibitemShut {NoStop}%
\bibitem [{\citenamefont {Rinnenthal}\ and\ \citenamefont
  {Gericke}(2002)}]{rinnenthal2002state}%
  \BibitemOpen
  \bibfield  {author} {\bibinfo {author} {\bibfnamefont {J.~L.}\ \bibnamefont
  {Rinnenthal}}\ and\ \bibinfo {author} {\bibfnamefont {K.-H.}\ \bibnamefont
  {Gericke}},\ }\href@noop {} {\bibfield  {journal} {\bibinfo  {journal} {J.
  Chem. Phys}\ }\textbf {\bibinfo {volume} {116}},\ \bibinfo {pages} {9776}
  (\bibinfo {year} {2002})}\BibitemShut {NoStop}%
\bibitem [{\citenamefont {Krems}\ \emph {et~al.}(2003)\citenamefont {Krems},
  \citenamefont {Sadeghpour}, \citenamefont {Dalgarno}, \citenamefont {Zgid},
  \citenamefont {K{\l}os},\ and\ \citenamefont
  {Cha{\l}asi{\'n}ski}}]{krems2003low}%
  \BibitemOpen
  \bibfield  {author} {\bibinfo {author} {\bibfnamefont {R.}~\bibnamefont
  {Krems}}, \bibinfo {author} {\bibfnamefont {H.}~\bibnamefont {Sadeghpour}},
  \bibinfo {author} {\bibfnamefont {A.}~\bibnamefont {Dalgarno}}, \bibinfo
  {author} {\bibfnamefont {D.}~\bibnamefont {Zgid}}, \bibinfo {author}
  {\bibfnamefont {J.}~\bibnamefont {K{\l}os}}, \ and\ \bibinfo {author}
  {\bibfnamefont {G.}~\bibnamefont {Cha{\l}asi{\'n}ski}},\ }\href@noop {}
  {\bibfield  {journal} {\bibinfo  {journal} {Phys. Rev. A}\ }\textbf {\bibinfo
  {volume} {68}},\ \bibinfo {pages} {051401} (\bibinfo {year}
  {2003})}\BibitemShut {NoStop}%
\bibitem [{\citenamefont {Cybulski}\ \emph {et~al.}(2005)\citenamefont
  {Cybulski}, \citenamefont {Krems}, \citenamefont {Sadeghpour}, \citenamefont
  {Dalgarno}, \citenamefont {K{\l}os}, \citenamefont {Groenenboom},
  \citenamefont {van~der Avoird}, \citenamefont {Zgid},\ and\ \citenamefont
  {Cha{\l}asi{\'n}ski}}]{cybulski2005interaction}%
  \BibitemOpen
  \bibfield  {author} {\bibinfo {author} {\bibfnamefont {H.}~\bibnamefont
  {Cybulski}}, \bibinfo {author} {\bibfnamefont {R.}~\bibnamefont {Krems}},
  \bibinfo {author} {\bibfnamefont {H.}~\bibnamefont {Sadeghpour}}, \bibinfo
  {author} {\bibfnamefont {A.}~\bibnamefont {Dalgarno}}, \bibinfo {author}
  {\bibfnamefont {J.}~\bibnamefont {K{\l}os}}, \bibinfo {author} {\bibfnamefont
  {G.}~\bibnamefont {Groenenboom}}, \bibinfo {author} {\bibfnamefont
  {A.}~\bibnamefont {van~der Avoird}}, \bibinfo {author} {\bibfnamefont
  {D.}~\bibnamefont {Zgid}}, \ and\ \bibinfo {author} {\bibfnamefont
  {G.}~\bibnamefont {Cha{\l}asi{\'n}ski}},\ }\href@noop {} {\bibfield
  {journal} {\bibinfo  {journal} {J. Chem. Phys}\ }\textbf {\bibinfo {volume}
  {122}},\ \bibinfo {pages} {094307} (\bibinfo {year} {2005})}\BibitemShut
  {NoStop}%
\bibitem [{\citenamefont {Stoecklin}(2009)}]{stoecklin2009combining}%
  \BibitemOpen
  \bibfield  {author} {\bibinfo {author} {\bibfnamefont {T.}~\bibnamefont
  {Stoecklin}},\ }\href@noop {} {\bibfield  {journal} {\bibinfo  {journal}
  {Phys. Rev. A}\ }\textbf {\bibinfo {volume} {80}},\ \bibinfo {pages} {012710}
  (\bibinfo {year} {2009})}\BibitemShut {NoStop}%
\bibitem [{\citenamefont {Tobo{\l}a}\ \emph {et~al.}(2011)\citenamefont
  {Tobo{\l}a}, \citenamefont {Dumouchel}, \citenamefont {K{\l}os},\ and\
  \citenamefont {Lique}}]{tobola2011calculations}%
  \BibitemOpen
  \bibfield  {author} {\bibinfo {author} {\bibfnamefont {R.}~\bibnamefont
  {Tobo{\l}a}}, \bibinfo {author} {\bibfnamefont {F.}~\bibnamefont
  {Dumouchel}}, \bibinfo {author} {\bibfnamefont {J.}~\bibnamefont {K{\l}os}},
  \ and\ \bibinfo {author} {\bibfnamefont {F.}~\bibnamefont {Lique}},\
  }\href@noop {} {\bibfield  {journal} {\bibinfo  {journal} {J. Chem. Phys}\
  }\textbf {\bibinfo {volume} {134}},\ \bibinfo {pages} {024305} (\bibinfo
  {year} {2011})}\BibitemShut {NoStop}%
\bibitem [{\citenamefont {Dumouchel}\ \emph {et~al.}(2012)\citenamefont
  {Dumouchel}, \citenamefont {K{\l}os}, \citenamefont {Tobo{\l}a},
  \citenamefont {Bacmann}, \citenamefont {Maret}, \citenamefont {Hily-Blant},
  \citenamefont {Faure},\ and\ \citenamefont {Lique}}]{dumouchel2012fine}%
  \BibitemOpen
  \bibfield  {author} {\bibinfo {author} {\bibfnamefont {F.}~\bibnamefont
  {Dumouchel}}, \bibinfo {author} {\bibfnamefont {J.}~\bibnamefont {K{\l}os}},
  \bibinfo {author} {\bibfnamefont {R.}~\bibnamefont {Tobo{\l}a}}, \bibinfo
  {author} {\bibfnamefont {A.}~\bibnamefont {Bacmann}}, \bibinfo {author}
  {\bibfnamefont {S.}~\bibnamefont {Maret}}, \bibinfo {author} {\bibfnamefont
  {P.}~\bibnamefont {Hily-Blant}}, \bibinfo {author} {\bibfnamefont
  {A.}~\bibnamefont {Faure}}, \ and\ \bibinfo {author} {\bibfnamefont
  {F.}~\bibnamefont {Lique}},\ }\href@noop {} {\bibfield  {journal} {\bibinfo
  {journal} {J. Chem. Phys}\ }\textbf {\bibinfo {volume} {137}},\ \bibinfo
  {pages} {114306} (\bibinfo {year} {2012})}\BibitemShut {NoStop}%
\bibitem [{\citenamefont {Ramachandran}\ \emph {et~al.}(2018)\citenamefont
  {Ramachandran}, \citenamefont {K{\l}os},\ and\ \citenamefont
  {Lique}}]{ramachandran2018new}%
  \BibitemOpen
  \bibfield  {author} {\bibinfo {author} {\bibfnamefont {R.}~\bibnamefont
  {Ramachandran}}, \bibinfo {author} {\bibfnamefont {J.}~\bibnamefont
  {K{\l}os}}, \ and\ \bibinfo {author} {\bibfnamefont {F.}~\bibnamefont
  {Lique}},\ }\href@noop {} {\bibfield  {journal} {\bibinfo  {journal} {J.
  Chem. Phys}\ }\textbf {\bibinfo {volume} {148}},\ \bibinfo {pages} {084311}
  (\bibinfo {year} {2018})}\BibitemShut {NoStop}%
\bibitem [{\citenamefont {Rinnenthal}\ and\ \citenamefont
  {Gericke}(2000)}]{rinnenthal2000state}%
  \BibitemOpen
  \bibfield  {author} {\bibinfo {author} {\bibfnamefont {J.~L.}\ \bibnamefont
  {Rinnenthal}}\ and\ \bibinfo {author} {\bibfnamefont {K.-H.}\ \bibnamefont
  {Gericke}},\ }\href@noop {} {\bibfield  {journal} {\bibinfo  {journal} {J.
  Chem. Phys}\ }\textbf {\bibinfo {volume} {113}},\ \bibinfo {pages} {6210}
  (\bibinfo {year} {2000})}\BibitemShut {NoStop}%
\bibitem [{\citenamefont {Kerenskaya}\ \emph {et~al.}(2005)\citenamefont
  {Kerenskaya}, \citenamefont {Schnupf}, \citenamefont {Heaven}, \citenamefont
  {van~der Avoird},\ and\ \citenamefont
  {Groenenboom}}]{kerenskaya2005experimental}%
  \BibitemOpen
  \bibfield  {author} {\bibinfo {author} {\bibfnamefont {G.}~\bibnamefont
  {Kerenskaya}}, \bibinfo {author} {\bibfnamefont {U.}~\bibnamefont {Schnupf}},
  \bibinfo {author} {\bibfnamefont {M.~C.}\ \bibnamefont {Heaven}}, \bibinfo
  {author} {\bibfnamefont {A.}~\bibnamefont {van~der Avoird}}, \ and\ \bibinfo
  {author} {\bibfnamefont {G.~C.}\ \bibnamefont {Groenenboom}},\ }\href@noop {}
  {\bibfield  {journal} {\bibinfo  {journal} {Phys. Chem. Chem. Phys.}\
  }\textbf {\bibinfo {volume} {7}},\ \bibinfo {pages} {846} (\bibinfo {year}
  {2005})}\BibitemShut {NoStop}%
\bibitem [{\citenamefont {Bouhafs}\ and\ \citenamefont
  {Lique}(2015)}]{bouhafs2015}%
  \BibitemOpen
  \bibfield  {author} {\bibinfo {author} {\bibfnamefont {N.}~\bibnamefont
  {Bouhafs}}\ and\ \bibinfo {author} {\bibfnamefont {F.}~\bibnamefont
  {Lique}},\ }\href@noop {} {\bibfield  {journal} {\bibinfo  {journal} {J.
  Chem. Phys}\ }\textbf {\bibinfo {volume} {143}},\ \bibinfo {pages} {184311}
  (\bibinfo {year} {2015})}\BibitemShut {NoStop}%
\bibitem [{\citenamefont {Dagdigian}(1989)}]{dagdigian1989}%
  \BibitemOpen
  \bibfield  {author} {\bibinfo {author} {\bibfnamefont {P.~J.}\ \bibnamefont
  {Dagdigian}},\ }\href@noop {} {\bibfield  {journal} {\bibinfo  {journal} {J.
  Chem. Phys}\ }\textbf {\bibinfo {volume} {90}},\ \bibinfo {pages} {6110}
  (\bibinfo {year} {1989})}\BibitemShut {NoStop}%
\bibitem [{\citenamefont {Kendall}\ \emph {et~al.}(1998)\citenamefont
  {Kendall}, \citenamefont {Cha{\l}asi{\'n}ski}, \citenamefont {K{\l}os},
  \citenamefont {Bukowski}, \citenamefont {Severson}, \citenamefont
  {Szczȩ{\'s}niak},\ and\ \citenamefont {Cybulski}}]{kendall1998}%
  \BibitemOpen
  \bibfield  {author} {\bibinfo {author} {\bibfnamefont {R.~A.}\ \bibnamefont
  {Kendall}}, \bibinfo {author} {\bibfnamefont {G.}~\bibnamefont
  {Cha{\l}asi{\'n}ski}}, \bibinfo {author} {\bibfnamefont {J.}~\bibnamefont
  {K{\l}os}}, \bibinfo {author} {\bibfnamefont {R.}~\bibnamefont {Bukowski}},
  \bibinfo {author} {\bibfnamefont {M.~W.}\ \bibnamefont {Severson}}, \bibinfo
  {author} {\bibfnamefont {M.}~\bibnamefont {Szczȩ{\'s}niak}}, \ and\ \bibinfo
  {author} {\bibfnamefont {S.~M.}\ \bibnamefont {Cybulski}},\ }\href@noop {}
  {\bibfield  {journal} {\bibinfo  {journal} {J. Chem. Phys}\ }\textbf
  {\bibinfo {volume} {108}},\ \bibinfo {pages} {3235} (\bibinfo {year}
  {1998})}\BibitemShut {NoStop}%
\bibitem [{\citenamefont {Cybulski}\ \emph {et~al.}(1995)\citenamefont
  {Cybulski}, \citenamefont {Burcl}, \citenamefont {Chal/asi{\'n}ski},\ and\
  \citenamefont {Szczȩ{\'s}niak}}]{cybulski1995partitioning}%
  \BibitemOpen
  \bibfield  {author} {\bibinfo {author} {\bibfnamefont {S.~M.}\ \bibnamefont
  {Cybulski}}, \bibinfo {author} {\bibfnamefont {R.}~\bibnamefont {Burcl}},
  \bibinfo {author} {\bibfnamefont {G.}~\bibnamefont {Chal/asi{\'n}ski}}, \
  and\ \bibinfo {author} {\bibfnamefont {M.}~\bibnamefont {Szczȩ{\'s}niak}},\
  }\href@noop {} {\bibfield  {journal} {\bibinfo  {journal} {J. Chem. Phys}\
  }\textbf {\bibinfo {volume} {103}},\ \bibinfo {pages} {10116} (\bibinfo
  {year} {1995})}\BibitemShut {NoStop}%
\bibitem [{\citenamefont {Cybulski}(1989)}]{cybulski1989trurl}%
  \BibitemOpen
  \bibfield  {author} {\bibinfo {author} {\bibfnamefont {S.}~\bibnamefont
  {Cybulski}},\ }\href@noop {} {\bibfield  {journal} {\bibinfo  {journal} {J.
  Chem. Phys}\ }\textbf {\bibinfo {volume} {91}},\ \bibinfo {pages} {7048}
  (\bibinfo {year} {1989})}\BibitemShut {NoStop}%
\bibitem [{\citenamefont {Kalugina}\ \emph {et~al.}(2014)\citenamefont
  {Kalugina}, \citenamefont {Lique},\ and\ \citenamefont
  {Marinakis}}]{kalugina2014new}%
  \BibitemOpen
  \bibfield  {author} {\bibinfo {author} {\bibfnamefont {Y.}~\bibnamefont
  {Kalugina}}, \bibinfo {author} {\bibfnamefont {F.}~\bibnamefont {Lique}}, \
  and\ \bibinfo {author} {\bibfnamefont {S.}~\bibnamefont {Marinakis}},\
  }\href@noop {} {\bibfield  {journal} {\bibinfo  {journal} {Phys. Chem. Chem.
  Phys.}\ }\textbf {\bibinfo {volume} {16}},\ \bibinfo {pages} {13500}
  (\bibinfo {year} {2014})}\BibitemShut {NoStop}%
\bibitem [{\citenamefont {Lique}(2015)}]{lique2015}%
  \BibitemOpen
  \bibfield  {author} {\bibinfo {author} {\bibfnamefont {F.}~\bibnamefont
  {Lique}},\ }\href@noop {} {\enquote {\bibinfo {title} {Communication:
  Rotational excitation of hcl by h: Rigid rotor vs. reactive approaches},}\ }
  (\bibinfo {year} {2015})\BibitemShut {NoStop}%
\bibitem [{\citenamefont {{Hampel}}\ \emph {et~al.}(1992)\citenamefont
  {{Hampel}}, \citenamefont {{Peterson}},\ and\ \citenamefont
  {{Werner}}}]{Hampel92}%
  \BibitemOpen
  \bibfield  {author} {\bibinfo {author} {\bibfnamefont {C.}~\bibnamefont
  {{Hampel}}}, \bibinfo {author} {\bibfnamefont {K.~A.}\ \bibnamefont
  {{Peterson}}}, \ and\ \bibinfo {author} {\bibfnamefont {H.-J.}\ \bibnamefont
  {{Werner}}},\ }\href {\doibase 10.1016/0009-2614(92)86093-W} {\bibfield
  {journal} {\bibinfo  {journal} {Chem. Phys. Lett.}\ }\textbf {\bibinfo
  {volume} {190}},\ \bibinfo {pages} {1} (\bibinfo {year} {1992})}\BibitemShut
  {NoStop}%
\bibitem [{\citenamefont {Watts}\ \emph {et~al.}(1993)\citenamefont {Watts},
  \citenamefont {Gauss},\ and\ \citenamefont {Bartlett}}]{watts1993coupled}%
  \BibitemOpen
  \bibfield  {author} {\bibinfo {author} {\bibfnamefont {J.~D.}\ \bibnamefont
  {Watts}}, \bibinfo {author} {\bibfnamefont {J.}~\bibnamefont {Gauss}}, \ and\
  \bibinfo {author} {\bibfnamefont {R.~J.}\ \bibnamefont {Bartlett}},\
  }\href@noop {} {\bibfield  {journal} {\bibinfo  {journal} {J. Chem. Phys}\
  }\textbf {\bibinfo {volume} {98}},\ \bibinfo {pages} {8718} (\bibinfo {year}
  {1993})}\BibitemShut {NoStop}%
\bibitem [{mol()}]{molpro10long}%
  \BibitemOpen
  \href@noop {} {}\bibinfo {note} {H.-J. Werner, P. J. Knowles, G. Knizia, F.
  R. Manby, M. {Sch\"{u}tz}, P. Celani, T. Korona, R. Lindh, A. Mitrushenkov,
  G. Rauhut, K. R. Shamasundar, T. B. Adler, R. D. Amos, A. Bernhardsson, A.
  Berning, D. L. Cooper, M. J. O. Deegan, A. J. Dobbyn, F. Eckert, E. Goll, C.
  Hampel, A. Hesselmann, G. Hetzer, T. Hrenar, G. Jansen, C. K\"{o}ppl, Y. Liu,
  A. W. Lloyd, R. A. Mata, A. J. May, S. J. McNicholas, W. Meyer, M. E. Mura,
  A. Nicklass, D. P. O'Neill, P. Palmieri, K. Pfl\"{u}ger, R. Pitzer, M.
  Reiher, T. Shiozaki, H. Stoll, A. J. Stone, R. Tarroni, T. Thorsteinsson, M.
  Wang and A. Wolf, \textit{MOLPRO}, \textit{version 2010.1}, a package of
  \textit{ab initio} programs, 2010, see http://www.molpro.net}\BibitemShut
  {NoStop}%
\bibitem [{\citenamefont {Boys}\ and\ \citenamefont
  {Bernardi}(1970)}]{boys1970calculation}%
  \BibitemOpen
  \bibfield  {author} {\bibinfo {author} {\bibfnamefont {S.~F.}\ \bibnamefont
  {Boys}}\ and\ \bibinfo {author} {\bibfnamefont {F.~d.}\ \bibnamefont
  {Bernardi}},\ }\href@noop {} {\bibfield  {journal} {\bibinfo  {journal} {Mol.
  Phys.}\ }\textbf {\bibinfo {volume} {19}},\ \bibinfo {pages} {553} (\bibinfo
  {year} {1970})}\BibitemShut {NoStop}%
\bibitem [{\citenamefont {Dunning~Jr}(1989)}]{dunning1989gaussian}%
  \BibitemOpen
  \bibfield  {author} {\bibinfo {author} {\bibfnamefont {T.~H.}\ \bibnamefont
  {Dunning~Jr}},\ }\href@noop {} {\bibfield  {journal} {\bibinfo  {journal} {J.
  Chem. Phys}\ }\textbf {\bibinfo {volume} {90}},\ \bibinfo {pages} {1007}
  (\bibinfo {year} {1989})}\BibitemShut {NoStop}%
\bibitem [{\citenamefont {Peterson}\ \emph {et~al.}(1994)\citenamefont
  {Peterson}, \citenamefont {Woon},\ and\ \citenamefont
  {Dunning~Jr}}]{peterson1994benchmark}%
  \BibitemOpen
  \bibfield  {author} {\bibinfo {author} {\bibfnamefont {K.~A.}\ \bibnamefont
  {Peterson}}, \bibinfo {author} {\bibfnamefont {D.~E.}\ \bibnamefont {Woon}},
  \ and\ \bibinfo {author} {\bibfnamefont {T.~H.}\ \bibnamefont {Dunning~Jr}},\
  }\href@noop {} {\bibfield  {journal} {\bibinfo  {journal} {J. Chem. Phys}\
  }\textbf {\bibinfo {volume} {100}},\ \bibinfo {pages} {7410} (\bibinfo {year}
  {1994})}\BibitemShut {NoStop}%
\bibitem [{\citenamefont {Werner}\ \emph {et~al.}(1988)\citenamefont {Werner},
  \citenamefont {Follmeg},\ and\ \citenamefont
  {Alexander}}]{werner1988adiabatic}%
  \BibitemOpen
  \bibfield  {author} {\bibinfo {author} {\bibfnamefont {H.-J.}\ \bibnamefont
  {Werner}}, \bibinfo {author} {\bibfnamefont {B.}~\bibnamefont {Follmeg}}, \
  and\ \bibinfo {author} {\bibfnamefont {M.~H.}\ \bibnamefont {Alexander}},\
  }\href@noop {} {\bibfield  {journal} {\bibinfo  {journal} {J. Chem. Phys}\
  }\textbf {\bibinfo {volume} {89}},\ \bibinfo {pages} {3139} (\bibinfo {year}
  {1988})}\BibitemShut {NoStop}%
\bibitem [{\citenamefont {Colbert}\ and\ \citenamefont
  {Miller}(1992)}]{colbert1992novel}%
  \BibitemOpen
  \bibfield  {author} {\bibinfo {author} {\bibfnamefont {D.~T.}\ \bibnamefont
  {Colbert}}\ and\ \bibinfo {author} {\bibfnamefont {W.~H.}\ \bibnamefont
  {Miller}},\ }\href@noop {} {\bibfield  {journal} {\bibinfo  {journal} {J.
  Chem. Phys}\ }\textbf {\bibinfo {volume} {96}},\ \bibinfo {pages} {1982}
  (\bibinfo {year} {1992})}\BibitemShut {NoStop}%
\bibitem [{\citenamefont {Werner}\ and\ \citenamefont
  {Knowles}(1988)}]{werner1988efficient}%
  \BibitemOpen
  \bibfield  {author} {\bibinfo {author} {\bibfnamefont {H.-J.}\ \bibnamefont
  {Werner}}\ and\ \bibinfo {author} {\bibfnamefont {P.~J.}\ \bibnamefont
  {Knowles}},\ }\href@noop {} {\bibfield  {journal} {\bibinfo  {journal} {J.
  Chem. Phys}\ }\textbf {\bibinfo {volume} {89}},\ \bibinfo {pages} {5803}
  (\bibinfo {year} {1988})}\BibitemShut {NoStop}%
\bibitem [{bou()}]{bound}%
  \BibitemOpen
  \href@noop {} {}\bibinfo {note} {J. M. Hutson, BOUND computer code, version 5
  (1993), distributed by Collaborative Computational Project No. 6 of the
  Science and Engineering Research Council (UK).}\BibitemShut {Stop}%
\bibitem [{\citenamefont {Jansen}\ \emph {et~al.}(1993)\citenamefont {Jansen},
  \citenamefont {Hess},\ and\ \citenamefont {Wormer}}]{jansen1993theoretical}%
  \BibitemOpen
  \bibfield  {author} {\bibinfo {author} {\bibfnamefont {G.}~\bibnamefont
  {Jansen}}, \bibinfo {author} {\bibfnamefont {B.~A.}\ \bibnamefont {Hess}}, \
  and\ \bibinfo {author} {\bibfnamefont {P.~E.}\ \bibnamefont {Wormer}},\
  }\href@noop {} {\bibfield  {journal} {\bibinfo  {journal} {Chem. Phys.
  Lett.}\ }\textbf {\bibinfo {volume} {214}},\ \bibinfo {pages} {103} (\bibinfo
  {year} {1993})}\BibitemShut {NoStop}%
\bibitem [{\citenamefont {Gordy}\ and\ \citenamefont {Cook}(1984)}]{gordy1984}%
  \BibitemOpen
  \bibfield  {author} {\bibinfo {author} {\bibfnamefont {W.}~\bibnamefont
  {Gordy}}\ and\ \bibinfo {author} {\bibfnamefont {R.~L.}\ \bibnamefont
  {Cook}},\ }\href@noop {} {\emph {\bibinfo {title} {Microwave molecular
  spectra}}}\ (\bibinfo  {publisher} {Wiley,},\ \bibinfo {year}
  {1984})\BibitemShut {NoStop}%
\bibitem [{\citenamefont {Lique}\ \emph {et~al.}(2005)\citenamefont {Lique},
  \citenamefont {Spielfiedel}, \citenamefont {Dubernet},\ and\ \citenamefont
  {Feautrier}}]{lique2005}%
  \BibitemOpen
  \bibfield  {author} {\bibinfo {author} {\bibfnamefont {F.}~\bibnamefont
  {Lique}}, \bibinfo {author} {\bibfnamefont {A.}~\bibnamefont {Spielfiedel}},
  \bibinfo {author} {\bibfnamefont {M.-L.}\ \bibnamefont {Dubernet}}, \ and\
  \bibinfo {author} {\bibfnamefont {N.}~\bibnamefont {Feautrier}},\ }\href@noop
  {} {\bibfield  {journal} {\bibinfo  {journal} {J. Chem. Phys}\ }\textbf
  {\bibinfo {volume} {123}},\ \bibinfo {pages} {134316} (\bibinfo {year}
  {2005})}\BibitemShut {NoStop}%
\bibitem [{\citenamefont {Hutson}\ and\ \citenamefont
  {Green}(1994)}]{molscat1994}%
  \BibitemOpen
  \bibfield  {author} {\bibinfo {author} {\bibfnamefont {J.~M.}\ \bibnamefont
  {Hutson}}\ and\ \bibinfo {author} {\bibfnamefont {S.}~\bibnamefont {Green}},\
  }\href@noop {} {} (\bibinfo {year} {1994}),\ \bibinfo {note} {{\sc molscat}
  computer code, version 14 (1994), distributed by Collaborative Computational
  Project No. 6 of the Engineering and Physical Sciences Research Council
  (UK)}\BibitemShut {NoStop}%
\bibitem [{\citenamefont {Lewen}\ \emph {et~al.}(2004)\citenamefont {Lewen},
  \citenamefont {Br{\"u}nken}, \citenamefont {Winnewisser}, \citenamefont
  {{\v{S}}ime{\v{c}}kov{\'a}},\ and\ \citenamefont {Urban}}]{lewen2004doppler}%
  \BibitemOpen
  \bibfield  {author} {\bibinfo {author} {\bibfnamefont {F.}~\bibnamefont
  {Lewen}}, \bibinfo {author} {\bibfnamefont {S.}~\bibnamefont {Br{\"u}nken}},
  \bibinfo {author} {\bibfnamefont {G.}~\bibnamefont {Winnewisser}}, \bibinfo
  {author} {\bibfnamefont {M.}~\bibnamefont {{\v{S}}ime{\v{c}}kov{\'a}}}, \
  and\ \bibinfo {author} {\bibfnamefont {{\v{S}}.}~\bibnamefont {Urban}},\
  }\href@noop {} {\bibfield  {journal} {\bibinfo  {journal} {Journal of
  Molecular Spectroscopy}\ }\textbf {\bibinfo {volume} {226}},\ \bibinfo
  {pages} {113} (\bibinfo {year} {2004})}\BibitemShut {NoStop}%
\bibitem [{\citenamefont {Smith}\ \emph {et~al.}(1979)\citenamefont {Smith},
  \citenamefont {Malik},\ and\ \citenamefont {Secrest}}]{smith1979rot}%
  \BibitemOpen
  \bibfield  {author} {\bibinfo {author} {\bibfnamefont {L.~N.}\ \bibnamefont
  {Smith}}, \bibinfo {author} {\bibfnamefont {D.~J.}\ \bibnamefont {Malik}}, \
  and\ \bibinfo {author} {\bibfnamefont {D.}~\bibnamefont {Secrest}},\
  }\href@noop {} {\bibfield  {journal} {\bibinfo  {journal} {J. Chem. Phys}\
  }\textbf {\bibinfo {volume} {71}},\ \bibinfo {pages} {4502} (\bibinfo {year}
  {1979})}\BibitemShut {NoStop}%
\bibitem [{\citenamefont {Christoffel}\ and\ \citenamefont
  {Bowman}(1983)}]{christoffel1983complex}%
  \BibitemOpen
  \bibfield  {author} {\bibinfo {author} {\bibfnamefont {K.~M.}\ \bibnamefont
  {Christoffel}}\ and\ \bibinfo {author} {\bibfnamefont {J.~M.}\ \bibnamefont
  {Bowman}},\ }\href@noop {} {\bibfield  {journal} {\bibinfo  {journal} {J.
  Chem. Phys}\ }\textbf {\bibinfo {volume} {78}},\ \bibinfo {pages} {3952}
  (\bibinfo {year} {1983})}\BibitemShut {NoStop}%
\bibitem [{\citenamefont {Alexander}\ and\ \citenamefont
  {Dagdigian}(1983)}]{alexander1983propensity}%
  \BibitemOpen
  \bibfield  {author} {\bibinfo {author} {\bibfnamefont {M.~H.}\ \bibnamefont
  {Alexander}}\ and\ \bibinfo {author} {\bibfnamefont {P.~J.}\ \bibnamefont
  {Dagdigian}},\ }\href@noop {} {\bibfield  {journal} {\bibinfo  {journal} {J.
  Chem. Phys}\ }\textbf {\bibinfo {volume} {79}},\ \bibinfo {pages} {302}
  (\bibinfo {year} {1983})}\BibitemShut {NoStop}%
\bibitem [{\citenamefont {Orlikowski}(1985)}]{orlikowski1985}%
  \BibitemOpen
  \bibfield  {author} {\bibinfo {author} {\bibfnamefont {T.}~\bibnamefont
  {Orlikowski}},\ }\href@noop {} {\bibfield  {journal} {\bibinfo  {journal}
  {Mol. Phys.}\ }\textbf {\bibinfo {volume} {56}},\ \bibinfo {pages} {35}
  (\bibinfo {year} {1985})}\BibitemShut {NoStop}%
\bibitem [{\citenamefont {Lique}(2010)}]{lique2010temperature}%
  \BibitemOpen
  \bibfield  {author} {\bibinfo {author} {\bibfnamefont {F.}~\bibnamefont
  {Lique}},\ }\href@noop {} {\bibfield  {journal} {\bibinfo  {journal} {J.
  Chem. Phys}\ }\textbf {\bibinfo {volume} {132}},\ \bibinfo {pages} {044311}
  (\bibinfo {year} {2010})}\BibitemShut {NoStop}%
\end{thebibliography}%

\end{document}